\documentclass[epsfig,useAMS,usegraphicx,usenatbib]{mn2e}
\usepackage{enumerate,amsmath,amssymb,fancybox,theorem,epsfig,graphicx,subfigure}
\def\Msun{\ifmmode{M_\odot}\else$M_\odot$\fi}
\def\msun{\ifmmode{M_\odot}\else$M_\odot$\fi}

\usepackage{amssymb}
\usepackage{lscape}
\title[Searching for AGNs among unidentified {\it INTEGRAL} sources]
{Searching for AGNs among unidentified {\it INTEGRAL} sources}

\author[E. Maiorano et al.]
{E. Maiorano$^{1}$\thanks{E-mail: \texttt{maiorano@iasfbo.inaf.it}}, 
R.~Landi$^{1}$, J.B. Stephen$^{1}$, L.~Bassani$^{1}$, N. Masetti$^{1}$, P. 
Parisi$^{1}$, \newauthor E. Palazzi$^{1}$, P. Parma$^{2}$, A.J. 
Bird$^{3}$, A. Bazzano$^{4}$, P. Ubertini$^{4}$, E. 
Jim\'enez-Bail\'on$^{5}$, \newauthor V. Chavushyan$^{6}$, G. Galaz$^{7}$, 
D. Minniti$^{7,8}$, L. Morelli$^{9}$\\
$^1$ IASF/INAF Bologna, via Piero Gobetti 101, 40129 Bologna, Italy\\
$^2$ IRA/INAF Bologna, via Piero Gobetti 101, 40129 Bologna, Italy\\
$^3$ School of Physics and Astronomy, University of Southampton, SO17 1BJ, 
	UK\\ 
$^4$ IASF/INAF Roma, via Fosso del Cavaliere 100, 00133 Rome, Italy\\
$^5$ Instituto de Astronom\'{\i}a, Universidad Nacional Aut\'onoma de 
	M\'exico, Apartado Postal 70-264, 04510 M\'exico D.F., M\'exico\\
$^6$ Instituto Nacional de Astrof\'{i}sica, \'Optica y Electr\'onica,
	Apartado Postal 51-216, 72000 Puebla, M\'exico\\
$^7$ Departamento de Astronom\'{i}a y Astrof\'{i}sica, Pontificia 
	Universidad Cat\'olica de Chile, Casilla 306, Santiago 22, Chile\\
$^8$ Specola Vaticana, V-00120 Citt\`a del Vaticano\\
$^9$ Dipartimento di Astronomia, Universit\`a di Padova, Vicolo 
	dell'Osservatorio 3, I-35122 Padua, Italy\\
}
\begin{document}

\date{}

\pagerange{\pageref{firstpage}--\pageref{lastpage}} \pubyear{2007}

\maketitle

\label{firstpage}

\begin{abstract}
We report on a new method to identify Active Galactic Nuclei (AGN) 
among unidentified {\it INTEGRAL} sources. This method consists of 
cross-correlating unidentified sources listed in the fourth IBIS Survey 
Catalogue first with infrared and then with radio catalogues and a 
posteriori verifying, by means of X-ray and optical follow up 
observations, the likelihood of these associations. In order to test this 
method, a sample of 8 sources has been extracted from the fourth IBIS 
Catalogue. For 7 sources of the sample we obtained an identification, 
whereas the last one (IGR J03103+5706) has insufficient information 
for a clear classification and deserves more in-depth study. We identified 
three objects (IGR J08190--3835, IGR J17520--6018, IGR J21441+4640) as 
AGNs and suggest that three more (IGR J00556+7708, IGRJ17219--1509, IGR 
J21268+6203) are likely active galaxies on the basis of their radio 
spectra, near-infrared photometry and location above the Galaxy plane. One 
source (IGR J05583--1257) has been classified as a starburst galaxy but it 
might have been spuriously associated with the {\it INTEGRAL} detection. 
\end{abstract}



\section{Introduction}

\begin{figure*}
\begin{center}
\centering{\mbox{\psfig{file=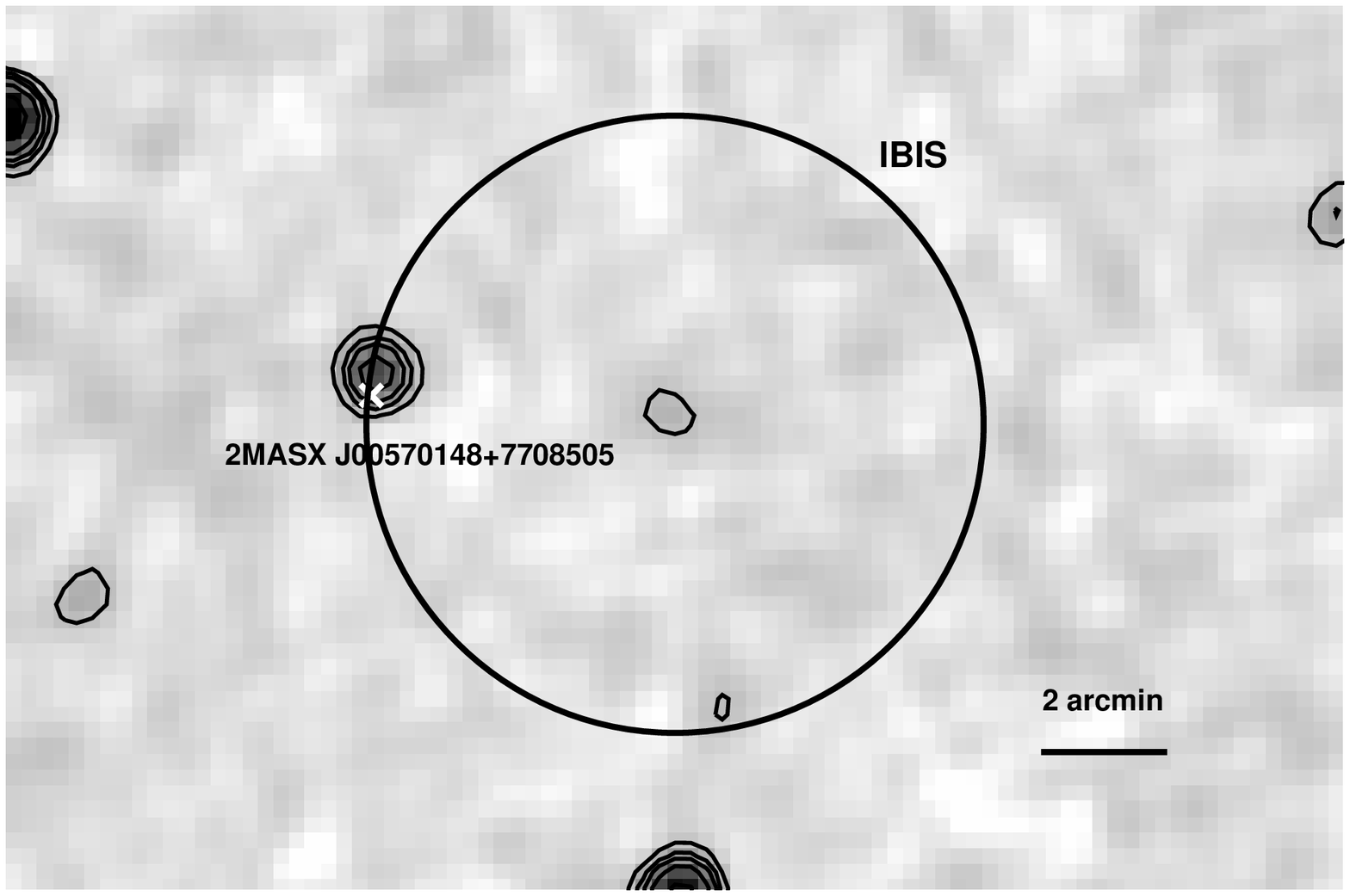,width=8cm}}}
\centering{\mbox{\psfig{file=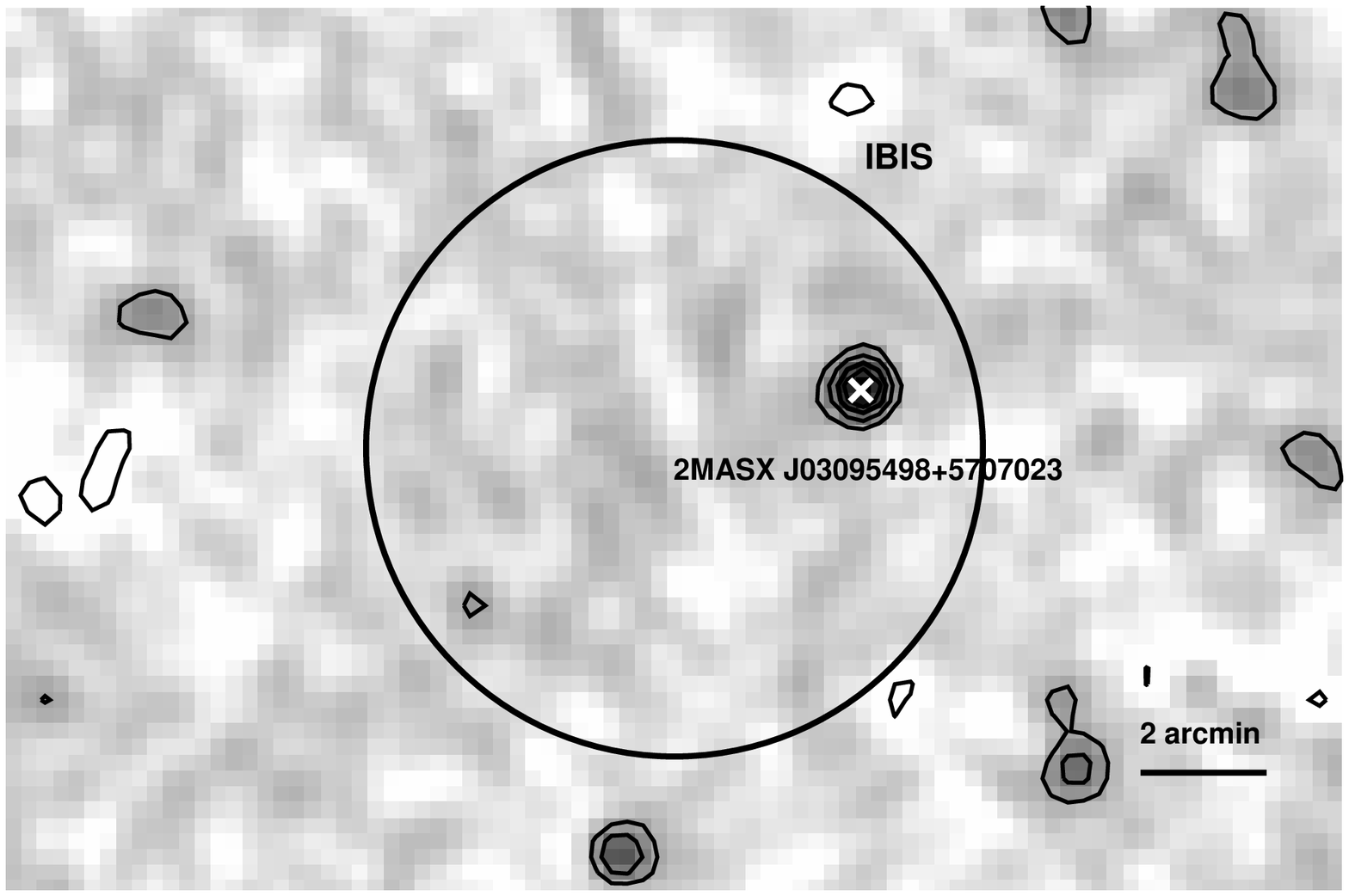,width=8cm}}}
\centering{\mbox{\psfig{file=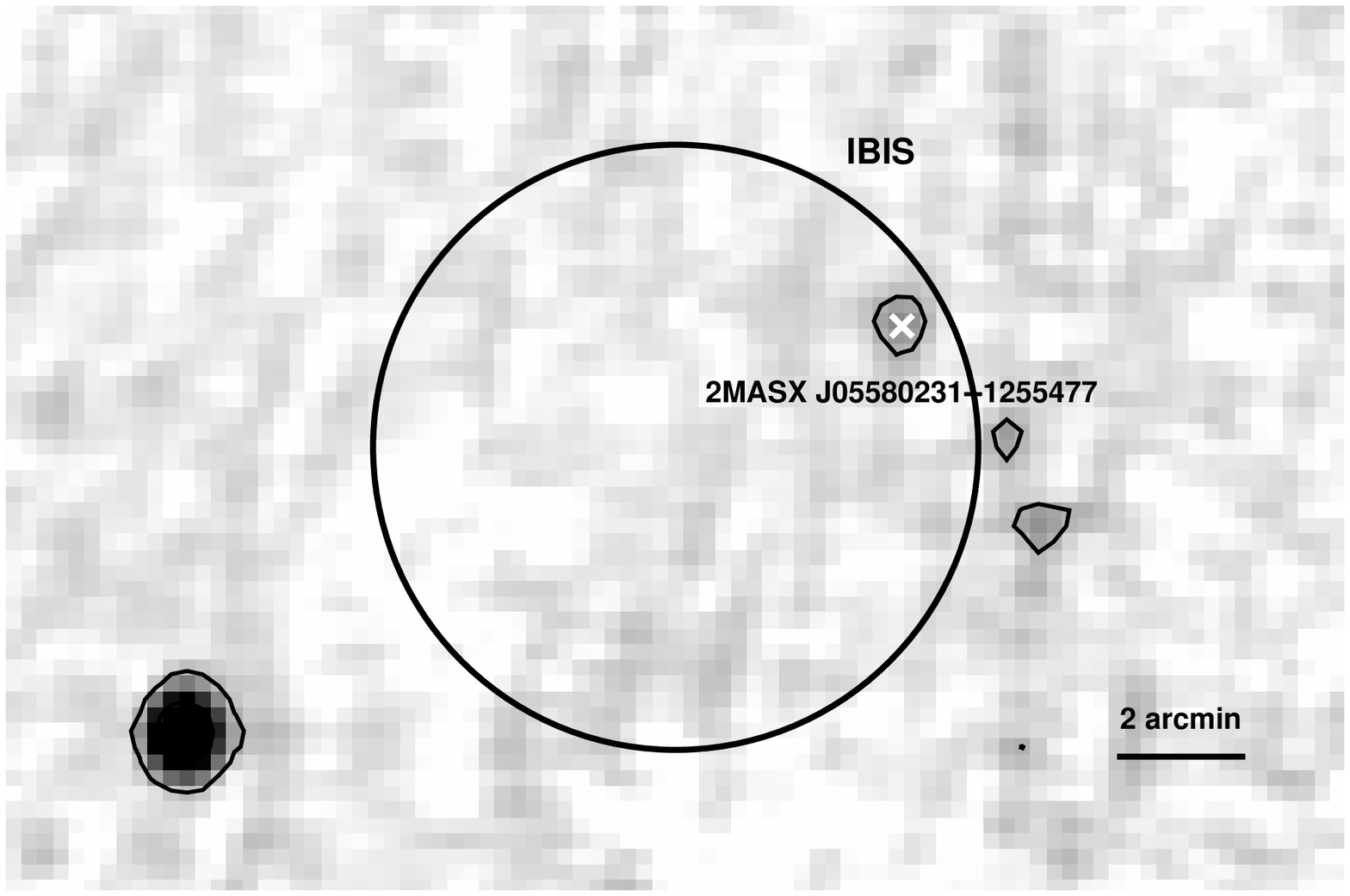,width=8cm}}}
\centering{\mbox{\psfig{file=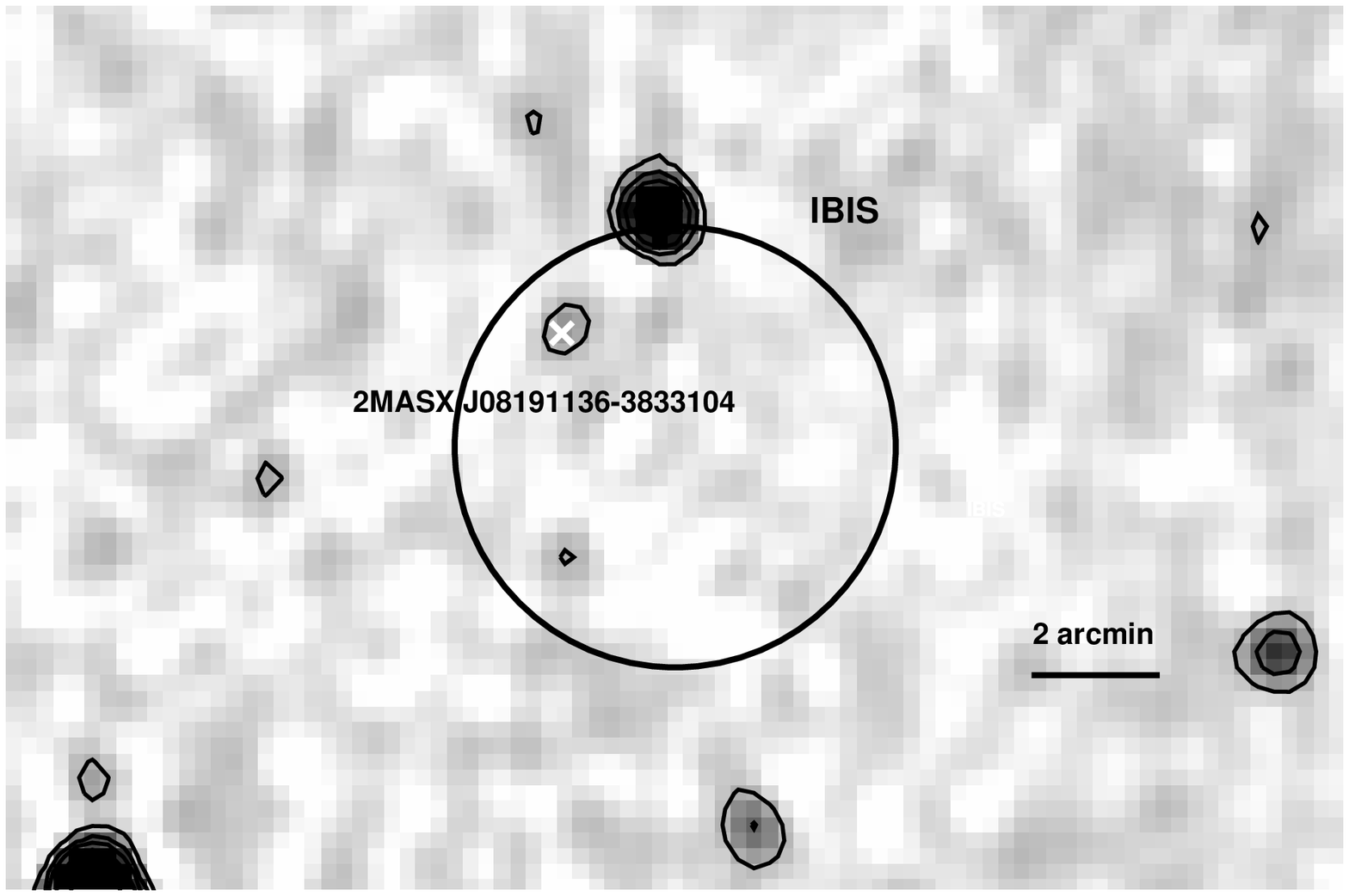,width=8cm}}}
\caption{NVSS image cut-outs for four sources in our sample with 
overimposed IBIS error circle and 2MASX source positions. In all images, 
the north is up and east to the left. The scale is reported at the 
bottom right corner.}
\end{center}
\end{figure*}

\begin{figure*}
\begin{center}
\centering{\mbox{\psfig{file=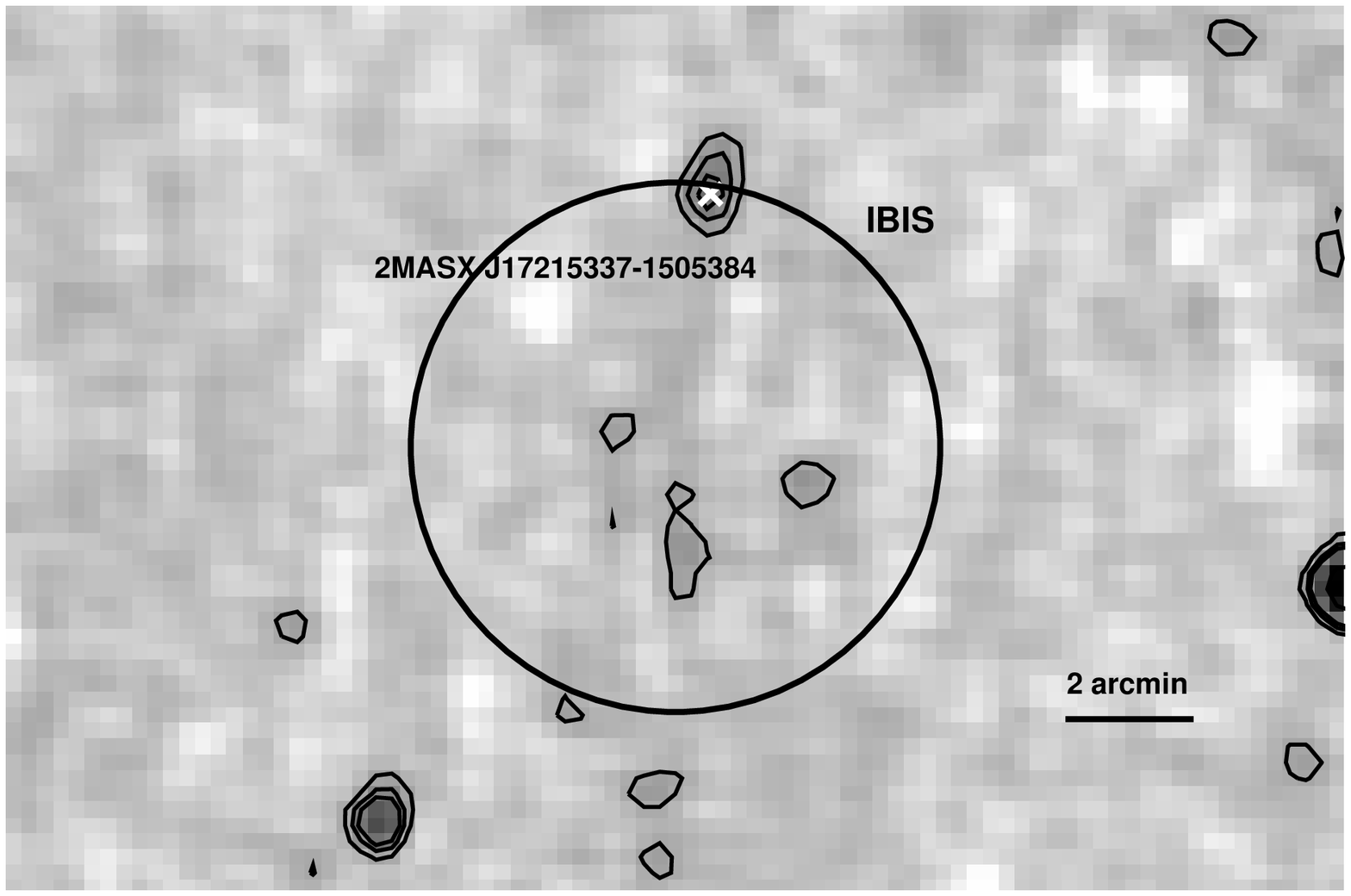,width=8cm}}}
\centering{\mbox{\psfig{file=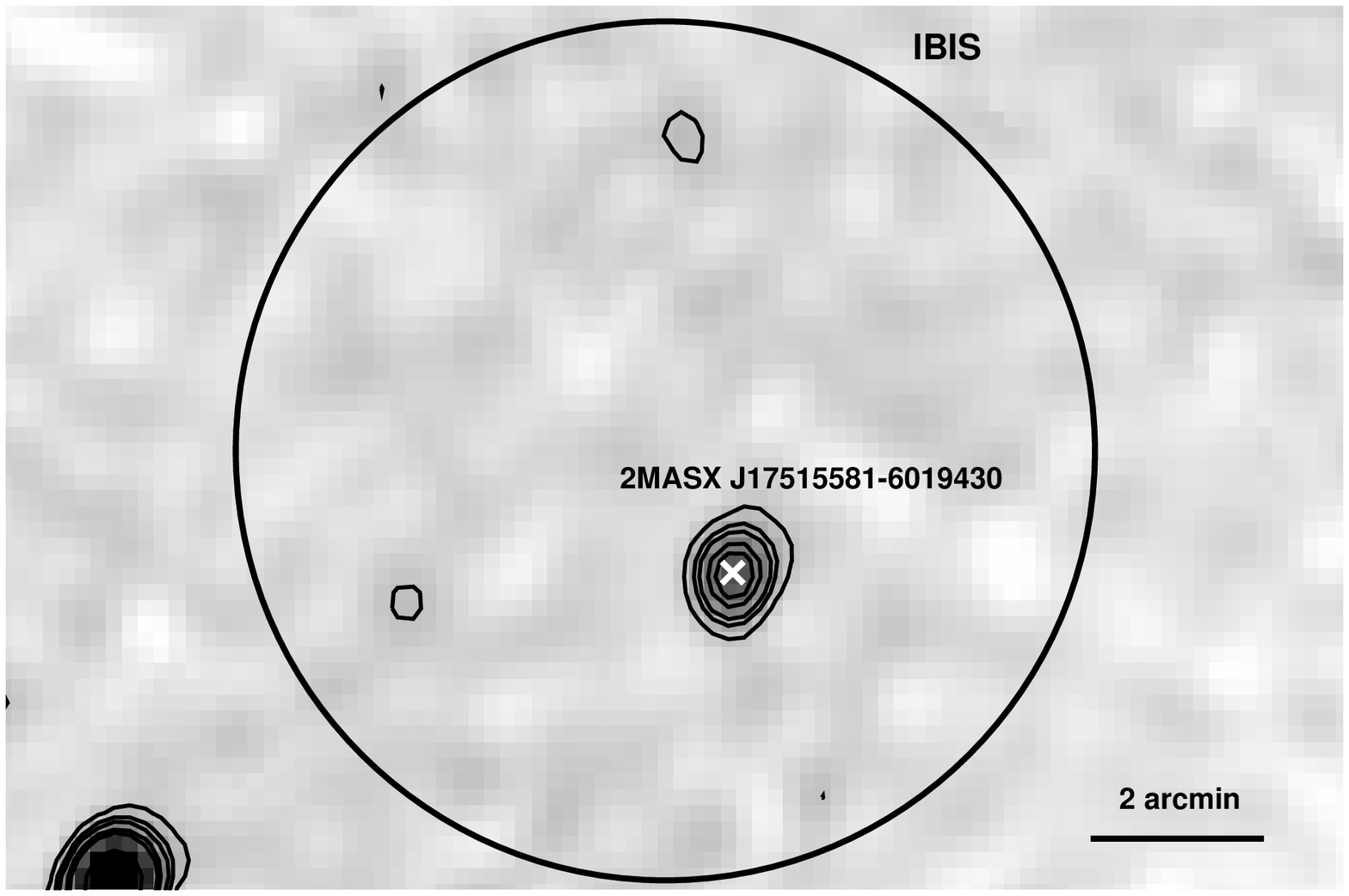,width=8cm}}}
\centering{\mbox{\psfig{file=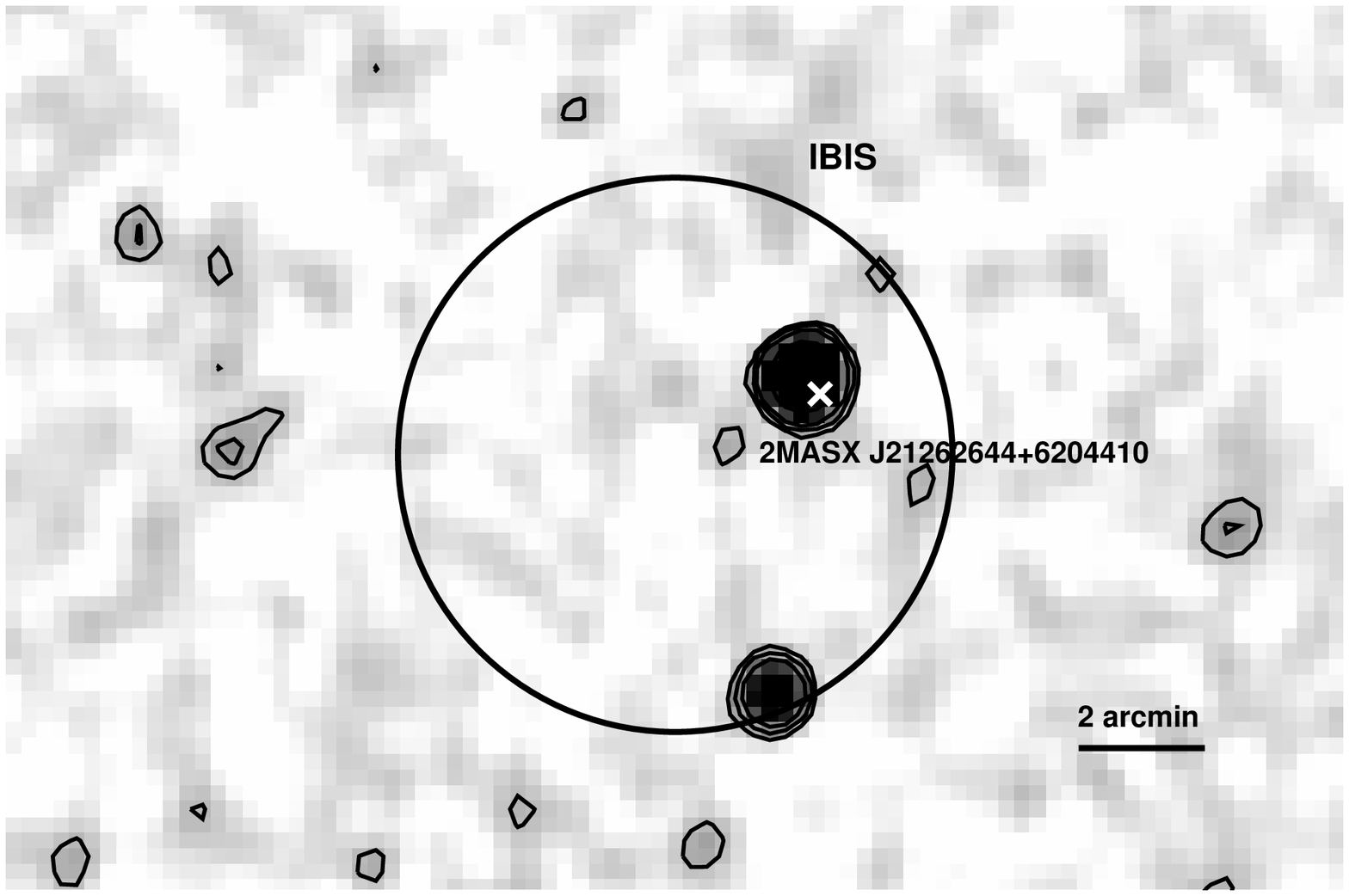,width=8cm}}}
\centering{\mbox{\psfig{file=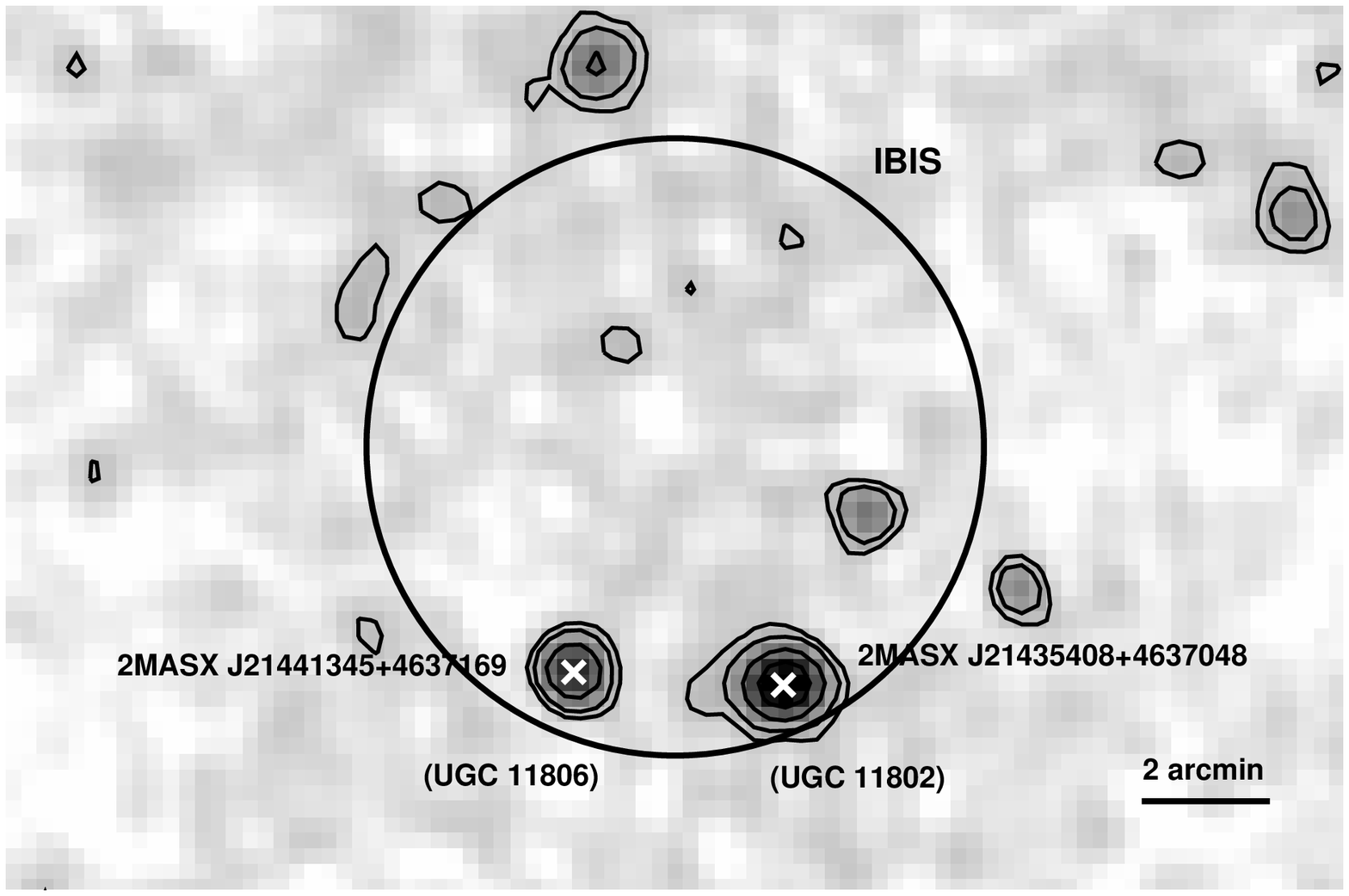,width=8cm}}}
\caption{NVSS or SUMSS image cut-outs for four sources in our sample with 
overimposed IBIS error circle and 2MASX source positions. In all images, 
the north is up and east to the left. The scale is reported at the 
bottom right corner.}
\end{center}
\end{figure*}

A key strategic objective of the {\it INTEGRAL} mission (Winkler et al. 
2003) is a survey of the sky at high energies ($>$ 20 keV), the domain 
where fundamental changes from primarily thermal to non-thermal 
sources/phenomena are expected, where the effects of absorption are 
drastically reduced and where most of the extreme astrophysical behaviour 
are taking place. To survey the high energy sky, {\it INTEGRAL} makes use 
of the unique imaging capability of the IBIS instrument (Ubertini et al. 
2003), which allows the detection of sources at the mCrab flux level with 
an angular resolution of 12$'$ and a point source location accuracy of 
typically 1--3$'$ within a large ($29\times29$ degrees) field of view.

So far, several surveys produced from data collected by IBIS have been 
reported in the literature, the most complete being that of Bird et al. 
(2010), which lists more than 700 sources of a diverse nature 
(Galactic and extra-galactic) and class. However, a large fraction 
($\sim$30\%) of these new {\it INTEGRAL} sources has no obvious 
counterpart in other wavebands and cannot be firmly classified; their 
classification is a primary objective of the survey work but it is made 
very difficult by the large IBIS error circles. Improved 
arcsecond-sized localization is therefore necessary to pinpoint the 
optical counterpart and through spectroscopic observations assess its 
nature/class (Masetti et al. 2010). For source identification one relies 
mostly on X-ray observations, but data in other wavebands can be used as 
well for counterpart search, in particular in those cases where the {\it 
INTEGRAL} unidentified source may be associated with an Active Galactic 
Nucleus (AGN).

Within this framework, in the present paper we present a new method for 
AGN identification in the 4th IBIS catalogue, which relies on 
cross-correlating unidentified sources first with infrared (IR) and then 
with radio catalogues. The method is a posteriori verified by means of 
X-ray and optical follow up observations of some of the sources discussed 
in the paper, which allow exploring their nature.

In the first step we use a set of extended IR objects (2MASS Extended 
Catalogue, Skrutskie et al. 2006) 97\% of which are associated with 
galaxies; we note that this is one of the most complete lists of galaxies 
available as it covers also the Galactic plane. Then we search for radio 
emission from these galaxies in order to identify likely AGN. This second 
step is justified by the fact that almost all AGN detected so far by {\it 
INTEGRAL} have a radio counterpart, which is not necessary radio-loud but 
can emit at a few mJy level.

An object which is extragalactic in nature and radio emitting, could also 
be a starburst galaxy but, in this case, we would not expect a bright 
X-ray luminosity as typically seen above 20 keV with currently operating 
hard X-ray detectors; indeed, so far, no starburst galaxy has been 
detected by IBIS in the 20--100 keV energy band.

Hence we use the IR, or better the information on source extension 
provided by the 2MASS extended catalogue, to pinpoint galaxies and then 
the information available from radio and hard X-ray emissions to look for 
AGN. We emphasize that many unidentified objects in the IBIS survey are on 
the Galactic plane, so that identification of these sources as AGN is 
not straightforward. Our method however provides a way to overcome 
problems related to the identification of active galaxies in the ``zone of 
avoidance''. Our choice of IR and radio catalogues is not intended to be a 
way to select special types of AGN, but rather a simple way to find 
galaxies among galactic and extragalactic sources and to select, among 
them, AGN associated with hard X-ray selected objects. X-ray follow up 
observations are instead used to provide confirmation of the proposed 
IR/radio and hard X-ray association while optical spectroscopy is 
performed to test the source AGN nature and class.

The paper is organized as follows: Sect. 2 describes the sample selection 
criteria and gives an overview of the extracted sample; the 
multiwavelength observations and analyses are reported in Sect. 3, while 
results and discussion are shown in Sect. 4. Finally, conclusions are 
presented in Sect. 5. The present work supersedes the analysis carried out 
in Maiorano et al. (2010), in which preliminary results for only three 
sources of the sample were presented.


\begin{table*}
\caption[]
{IBIS, 2MASX and radio positions (and corresponding errors) for each 
source in the sample. The radio positions come from the analysis 
carried out in this work (Subsect. 2.1).}
\scriptsize
\begin{center}
\begin{tabular}{ccccccc}
\noalign{\smallskip}
\hline
\hline
\noalign{\smallskip}
 Source & IBIS Position & IBIS & 2MASX Object & 2MASX Position & 
Radio Object & Radio Position \\
  & RA (J2000) & error circle &  & RA (J2000) &  & RA (J2000)(err) \\
  & Dec (J2000) & (arcmin) &  & Dec (J2000) &  & Dec (J2000)(err) \\
\noalign{\smallskip}
\hline
\hline
\noalign{\smallskip}

IGR J00556+7708 & 00:55:34.8 & 4.9$'$ & 2MASX J00570148+7708505 & 
00:57:01.482 & NVSS J005700+770911 & 00:57:00.016 (0.181s)\\
 & +77:08:24 &  &  & +77:08:50.50 &  & +77:09:11.81 (0.62$''$)\\

IGR J03103+5706 & 03:10:16.8 & 4.9$'$ & 2MASX J03095498+5707023 &  
 03:09:54.987 & NVSS J030954+570704 & 03:09:54.955 (0.128s)\\
  & +57:06:07.2 &  &  & +57:07:02.34 &  & +57:07:04.29 (1.04$''$)\\

IGR J05583--1257 & 05:58:17.04 & 4.8$'$ & 2MASX J05580231--1255477 &
 05:58:02.313 & NVSS J055802--125545 & 05:58:02.416 (0.172s)\\
  & --12:57:43.2 &  & (LCSB 0289O) & --12:55:47.78 &  & --12:55:45.54 
(3.30$''$)\\

IGR J08190--3835 & 08:19:02.16 & 3.5$'$ & 2MASX J08191136--3833104 &
 08:19:11.365 & NVSS J081910--383307 & 08:19:10.929 (0.219s)\\
  & --38:34:58.8 &  &  & --38:33:10.46 &  & --38:33:06.78 (3.00$''$)\\

IGR J17219--1509 & 17:21:55.68 & 4.2$'$ & 2MASX J17215337--1505384 &
 17:21:53.379 & NVSS J172153--150532 & 17:21:53.273 (0.155s)\\
  & --15:09:39.6 &  &  & --15:05:38.49 &  & --15:05:32.02 (3.70$''$)\\

IGR J17520--6018 & 17:52:02.16 & 5.0$'$ & 2MASX J17515581--6019430 &
 17:51:55.818 & SUMSS J175155--601943 & 17:51:55.585 (0.142s)\\
  & --60:18:18 &  &  & --60:19:43.08 &  & --60:19:43.87 (1.30$''$)\\

IGR J21268+6203 & 21:26:46.08 & 4.4$'$ & 2MASX J21262644+6204410 &
 21:26:26.440 & NVSS J212628+620457 & 21:26:28.707 (0.027s)\\
  & +62:03:43.2 &  &  & +62:04:41.03 &  & +62:04:57.58 (0.19$''$)\\

IGR J21441+4640 & 21:44:04.08 & 4.9$'$ & 2MASX J21441345+4637169 &
 21:44:13.455 & NVSS J214413+463718 & 21:44:13.419 (0.092s)\\
 & +46:40:51.6 &  & (UGC 11806) & +46:37:16.97 &  & +46:37:17.79 
(0.98$''$)\\
  &  &  & 2MASX J21435408+4637048 & 21:43:54.082 & NVSS J214354+463705 & 
21:43:54.055 (0.074s)\\
  &  &  & (UGC 11802) & +46:37:04.84 &  & +46:37:04.99 (0.66$''$)\\

\noalign{\smallskip}
\hline
\hline
\noalign{\smallskip}
\end{tabular}
\end{center}
\end{table*}


\section{Sample selection}

Historically, AGNs were discovered with radio observations, i.e. the radio 
selection is often a way to recognize active galaxies, except at lower 
luminosities where star-formation in galaxies can produce radio emission. 
Therefore, for bright objects, a mere detection in radio provides support 
for the presence of an active galaxy. Sample contamination from Galactic 
sources may however come from pulsars, microquasars and Cataclismic 
Variables (CVs). In many cases, association with a galaxy via 
cross-correlation with galaxy catalogues can help in selecting only 
extragalactic objects and, by means of the radio detection, pinpointing 
those sources that are likely AGNs.

So, while a mere radio detection does not imply an identification with an 
AGN, its combination with high energy X/gamma-ray emission, together with 
the association with a galaxy, strongly argues in favour of the 
identification of an unclassified {\it INTEGRAL} source with an active 
galaxy. Following this reasoning we have cross-correlated our set of 
unidentified {\it INTEGRAL} sources in the 4th IBIS catalogue (Bird 
et al. 2010) with IR/radio catalogues, in order to extract a small 
sample of objects likely associated with an AGN.

For the IR bands we have used the Two Micron All Sky Survey Extended 
(2MASX) Source Catalog (Skrutskie et al. 2006) which is a powerful tool to 
identify, within the sample of unidentified {\it INTEGRAL} sources, those 
possibly associated with galaxies. For this survey, the entire sky was 
uniformly scanned in three near-infrared (NIR) bands (J,H,K) to detect and 
characterise sources brighter than about 1 mJy in each band, with a 
signal-to-noise ratio greater than 10 and which are resolved and extended 
beyond the Two Micron All Sky Survey (2MASS) beam/point spread function. 
The absolute astrometric accuracy of the 2MASX catalogue is better than 
one arcsec. The extended source catalogue consists of 1,647,599 objects, 
97\% of which are galaxies while the remaining $\sim$3\% is made of 
sources in the Milky Way (mostly double and triple stars, HII regions, 
planetary and reflection nebulae), which are not expected to emit at high 
energies. Therefore, the presence of a 2MASX object inside the IBIS error 
circle suggests an association with a galaxy.

As radio catalogues we have used the NRAO VLA Sky Survey (NVSS, Condon et 
al. 1998) and the Sydney University Molonglo Sky Survey (SUMSS, Mauch et 
al. 2003) which are particularly well suited for finding counterparts of 
unidentified {\it INTEGRAL} sources: they are similar in sensitivity and 
spatial resolution and together they cover the whole sky. The NVSS 
catalogue covers the sky north of the J2000.0 Declination of --40 degrees 
(82\% of the celestial sphere) at 1.4 GHz (20 cm). The catalogue consists 
of almost 2 million discrete sources stronger than a flux density of about 
2.5 mJy. The NVSS images have 45 arcsecond FWHM angular resolution and 
nearly uniform sensitivity. The {\it rms} uncertainties in right ascension 
and declination vary from about 1$''$ for the 400,000 sources stronger 
than 15 mJy to 7$''$ at the survey limit. The SUMSS catalogue covers 
instead the sky south of the J2000.0 Declination of --30 degrees 
($\sim$20\% of the celestial sphere) and is carried out at 843 MHz (36 
cm). The survey consists of $4.3^{\circ} \times 4.3^{\circ}$ mosaic images 
with a resolution of $45\times45$ cosec$|\delta|$ arcsec$^{2}$, and a {\it 
rms} noise level of 1-2 mJy/beam. Positions in the catalogue are accurate 
to within 1-2$''$ for sources with peak brightness $>$20 mJy/beam, and are 
always better than 10$'$. The internal flux density scale is accurate to 
within 3\%. The radio detection of a 2MASX emitting galaxy strongly 
indicates that the source is an AGN if also detected above 10 keV.

To perform the correlation, we used the standard statistical technique 
which has been employed very successfully in other cases (Stephen et al. 
2005, 2006, 2010). This consists of simply calculating the number of {\it 
INTEGRAL} sources for which at least one 2MASX counterpart was within a 
specified distance, out to a distance where all {\it INTEGRAL} sources had 
at least one NIR counterpart. To have a control group we created a list of 
fake ``anti-{\it INTEGRAL}'' sources. For every object in the {\it 
INTEGRAL} list, we made a corresponding source in the fake list with 
coordinates mirrored in Galactic longitude and latitude (this mirroring 
was chosen due to the strong Galactic component evident in the {\it 
INTEGRAL} distribution), and the same correlation algorithm was then 
applied between this list and the 2MASX catalogue. Subtracting from the 
number of correlations in the true list those obtained in the false 
sample, it is possible to estimate the number of true associations. We see 
that the radius at which the first correlations between the ``anti-{\it 
INTEGRAL}'' sources sample and 2MASX catalogue appear is about 6 arcmin. 
This is comparable with the size of the IBIS error circle radius, 
therefore we expect at most 1 spurious correlation within the lot of 
selected sources (see below). The sample of associations extracted in this 
way, i.e. a list of objects likely associated with galaxies, was then 
cross-correlated to the radio catalogues following the same method.

By means of this sequence of cross-correlations we extracted a final 
sample of 8 objects which are seen in all 3 wavebands (hard X-rays, NIR 
and radio); all 8 can be considered as AGN candidates because they are 
classified as galaxies in the 2MASX and are detected both in radio and 
hard X-rays. Note that in one case we have two radio/NIR objects 
associated with a unique {\it INTEGRAL} source (IGR J21441+4640); both are 
detected in the 2MASX and radio catalogues and so are equally possible 
counterparts of the {\it INTEGRAL} source and as such will be considered 
in the following sections. In NASA/IPAC Extragalactic Database (NED) these 
two objects form a galaxy pair.

\subsection{Overview of the extracted sample} 
 
All objects in the sample are reported in Table 1; for each source we list 
the {\it INTEGRAL} name, the IBIS position with relative error, the IR 
(2MASX) and radio (NVSS or SUMSS) positions and associated errors of the 
putative counterparts. In our sample, 7 sources have a counterpart in the 
NVSS catalogue while only one has an association in the SUMSS.

The 2MASX catalogue provides J, H, K magnitudes for all sources in the 
sample which are reported in Table 2; note that all objects listed in 
Table 2 are classified as galaxies in NED. In Table 2 we also show the NIR 
colour indices J--H and H--K; these can be used as a tool to confirm the 
AGN nature of our sources. Indeed, all objects but one (2MASX 
J03095498+5707023) have NIR colours compatible with those of nearby active 
galaxies (see Fig. 1 of Kouzuma \& Yamaoka, 2010). Despite the 
uncertainties introduced by this method (i.e. it is not obvious from the 
work of Kouzuma \& Yamaoka [2010], which is the efficiency of finding an 
AGN via IR photometry), it is reassuring that almost all of our 
objects are compatible with NIR AGN colour indices.

Hard X-ray information concerning the sources of our sample is reported in 
Bird et al. (2010). We note here that all but two objects are classified 
as variable in the 4th IBIS catalogue since they are detected at the level 
of Revolution, Revolution Sequence or through the bursticity analysis; the 
only exceptions are IGR J05583--1257 and IGR J08190--3835 which are 
reported as persistent objects in the IBIS survey. All but 3 objects (IGR 
J03103+5706, IGR J08190-3835, IGR J21441+4640) are at Galactic latitude 
$>10^{\circ}$ an additional argument in favour of their extragalactic 
nature.

In order to provide radio images and fluxes, we have used the standard 
procedure employed within the software package 
AIPS\footnote{http://www.aips.nrao.edu/}, release 31DEC10 (Astronomical 
Image Processing System). NVSS/SUMSS fluxes extracted from the radio maps 
are reported in Table 3.

Figures 1 and 2 show the collection of NVSS/SUMSS image cut-outs for all 
of our sources with overimposed IBIS error circle and 2MASX source 
positions. For two sources (IGR J00556+7708, IGR J21268+6203) in our 
sample the radio and NIR positions are significantly different from each 
other, being separated by 22$''$ and 24$''$, respectively. In both cases 
we searched among the optical and NIR catalogues selected in the Vizier 
Catalogues Selection 
Page\footnote{http://vizier.u-strasbg.fr/viz-bin/VizieR} (USNO-B1, Monet 
et al. 2003, 
USNO-A2.0\footnote{http://www.nofs.navy.mil/projects/pmm/USNOSA2doc.html}, 
2MASS), but we could not find any optical or NIR counterparts within the 
radio error circle; furthermore, we note that the 2MASX galaxies lie in 
each case within the edge of the radio contours and are obviously extended 
in the NIR band (semimajor axis of 16.6$''$ and 13.7$''$, respectively). 
Indeed, in both cases the 2MASX and radio objects are associated with each 
other in NED on the basis of a sophisticated cross-identification analysis 
which takes into account not only the positional uncertainty but also the 
source extension (Mazzarella et al. 2001). Taking all these evidences into 
account, we decide to keep both {\it INTEGRAL} sources in the sample also 
in view of the fact that either the NIR or the radio object (or in the 
best case both) are likely counterparts to the high energy emitter.




By looking for further radio information in the data archives of different 
radio telescopes and also in the literature, we found that some sources in 
the sample have been observed at more than one radio frequency. In these 
few cases we have calculated the radio spectral index using the available 
data points and the usual relation F$_\nu \propto \nu^{-\alpha}$ (see 
Sect. 3). Furthermore, when the redshift is available, we have calculated 
the radio power at 1.4 GHz; all these values are reported in Table 3. Note 
that all objects in Table 3 have compact morphology in radio.


\begin{table*}
\caption[]{IR magnitudes as quoted in the 2MASS extended catalogue for all 
objects in the sample. In the case of IGR J21441+4640, we report the IR 
magnitudes and colours of both galaxies in the pair individually.}
\begin{center}
\begin{tabular}{cccccc}
\noalign{\smallskip}
\hline
\hline
\noalign{\smallskip}
 Source & J-mag & H-mag & K-mag & J--H & H--K \\
\noalign{\smallskip}
\hline
\hline
\noalign{\smallskip}

2MASX J00570148+7708505 & 14.922$\pm$0.222 & 14.291$\pm$0.333 & 
13.270$\pm$0.173 & 0.63 & 1.02 \\

2MASX J03095498+5707023 & 15.497$\pm$0.301 & 13.831$\pm$0.143 & 
12.772$\pm$0.096 & 1.67 & 1.06 \\

2MASX J05580231--1255477 & 13.192$\pm$0.080 & 12.608$\pm$0.119 & 
12.111$\pm$0.133 & 0.58 & 0.50 \\ 

2MASX J08191136--3833104 & 12.529$\pm$0.058 & 11.303$\pm$0.037 & 
10.818$\pm$0.049 & 1.23 & 0.49 \\ 

2MASX J17215337--1505384 & 15.563$\pm$0.237 & $>$14.520 & 13.985$\pm$0.181 
& 1.04 & 0.53 \\

2MASX J17515581--6019430 & 14.525$\pm$0.133 & 13.564$\pm$0.126 & 
13.408$\pm$0.166 & 0.96 & 0.15 \\

2MASX J21262644+6204410 & 14.777$\pm$0.185 & 14.172$\pm$0.234 & 
13.314$\pm$0.174 & 0.6 & 0.86 \\

2MASX J21441345+4637169 & 12.106$\pm$0.051 & 11.530$\pm$0.062 & 
11.312$\pm$0.074 & 0.57 & 0.22 \\
(UGC 11806) &  &  & \\

2MASX J21435408+4637048 & 11.814$\pm$0.055 & 11.088$\pm$0.060 & 
10.743$\pm$0.064 & 0.73 & 0.34 \\
(UGC 11802) &  &  & \\

\noalign{\smallskip}
\hline
\hline
\noalign{\smallskip}
\end{tabular}
\end{center}
\end{table*}


\begin{table*}
\caption[]{Radio information of all objects in the sample. NVSS 1.4 GHz 
(20 cm) and SUMSS 843 MHz (36 cm) fluxes are from this work. Other flux 
values are from the Westerbork Northern Sky Survey (WENSS) at 92 cm and 
the Parkes MIT-NRAO (PMN) Surveys at 6 cm. All the redshifts reported in 
the Table are from NED, except for NVSS J081910--383307 (2MASX 
J08191136--3833104) that is derived from the spectroscopic data analysis 
carried out in this work (Sect. 3).}
\begin{center}
\begin{tabular}{cccccccc}
\noalign{\smallskip}
\hline
\hline
\noalign{\smallskip}
 Source & Redshift & 325 MHz & 843 MHz & 1400 MHz & 4850 MHz & Spectral 
Index & Radio Power \\
  &  & 92 cm & $\sim$ 36 cm & 20 cm & $\sim$ 6 cm & $\alpha_{325{\rm 
MHz}}^{1.4{\rm GHz}}$ & at 1.4 GHz \\
  &  & (mJy) & (mJy) & (mJy) & (mJy) & (F$_\nu \propto \nu^{-\alpha}$) & 
(WHz$^{-1}$) \\
\noalign{\smallskip}
\hline
\hline
\noalign{\smallskip}

NVSS J005700+770911 &  & 51$\pm$2.1 &  & 16.0$\pm$1.2 &  & 
0.79$^{+0.09}_{-0.08}$ & \\ 

NVSS J030954+570704 &  & 21$\pm$3.8 &  & 10.2$\pm$1.0 &  & 
0.50$^{+0.18}_{-0.20}$ & \\

NVSS J055802--125545 & 0.003042 &  &  & 4.7$\pm$0.8 &  &  & 
9.31$\times10^{19}$ \\

NVSS J081910--383307 & 0.009 &  &  & 2.7$\pm$0.7 &  &  & 
4.71$\times10^{20}$ \\

NVSS J172153--150532 &  &  &  & 5.2$\pm$1.0 &  &  & \\

SUMSS J175155--601943 &  &  & 25.2$\pm$2.2 &  &  &  & \\

NVSS J212628+620457 &  & 29$\pm$4.8 &  & 45.1$\pm$1.4 & 19$\pm$4 & 
--0.30$^{+0.13}_{-0.16}$ & \\

NVSS J214413+463718 & 0.011081 & 23$\pm$4 &  & 11.0$\pm$1.1 &  & 
0.51$^{+0.17}_{-0.19}$ & 2.92$\times10^{21}$ \\
(UGC 11806) &  &  &  &  & \\

NVSS J214354+463705 & 0.010517 & 40$\pm$4 &  & 22.9$\pm$1.2 &  & 
0.38$^{+0.09}_{-0.10}$ & 5.47$\times10^{21}$\\
(UGC 11802) &  &  &  &  & \\

\noalign{\smallskip}
\hline
\hline
\noalign{\smallskip}
\end{tabular}
\end{center}
\end{table*}


\section{Follow up observations}

In order to test the validity of our method as well as to confirm the AGN 
nature of our objects, we have obtained a set of X-ray and optical 
follow up observations. The X-ray observations, carried out with 
Swift/XRT\footnote{The Swift/XRT observations were performed in the 
context of the approved follow-up program in collaboration with the Swift 
team.}, were useful to confirm the association between the NIR/radio 
source and the IBIS hard X-ray emission and hence to pinpoint the optical 
counterpart; while optical spectra of this counterpart allowed us to 
assess the source AGN nature and class.


\begin{table*}
\caption[]{Spectral parameters derived from the X-ray data analysis of 
four sources observed by Swift/XRT. The values of the Galactic column 
densities are from Kalberla et al. (2005).}
\begin{center}
\begin{tabular}{ccccccc}
\noalign{\smallskip}
\hline
\hline
\noalign{\smallskip}
 Source & Exposure & Count rate & $\Gamma$ & N$_{\rm H}(Gal)$ & N$_{\rm 
H}(intr)$ & Flux(2-10 keV) \\ 
  & (s) & (counts/s) & Photon index & (cm$^{-2}$) & (cm$^{-2}$) & (erg 
cm$^{-2}$s$^{-1}$) \\
\noalign{\smallskip}
\hline
\hline
\noalign{\smallskip}



IGR J05583--1257 &  &  &  &  &  & \\

(LCSB 0289O) & 4623 & $<$0.7$\times$10$^{-3}$ & 1.8 (fixed) & 
1.63$\times$10$^{21}$ &  & $<$5.8$\times$10$^{-14}$ \\

IGR J08190--3835 & 10310 & (18.6$\pm$1.3)$\times$10$^{-3}$ & 1.8 (fixed) & 
9.6$\times$10$^{21}$ & 13.6$\times$10$^{22}$ & 1.49$\times$10$^{-12}$ \\


IGR J17520--6018 & 11883 & (18.2$\pm$1.2)$\times$10$^{-3}$ & 1.8 (fixed) & 
7.0$\times$10$^{20}$ & 13$\times$10$^{22}$ & 2.55$\times$10$^{-12}$\\

 
IGR J21441+4640 & 2755 &  &  &  &  & \\

(UGC 11806) &  & (3.3$\pm$1.2)$\times$10$^{-3}$ & 1.8 (fixed) & 
2.57$\times$10$^{21}$ &  & 1.2$\times$10$^{-13}$ \\

(UGC 11802) &  & $<$1.2$\times$10$^{-3}$ & 1.8 (fixed) & 
2.8$\times$10$^{21}$ &  & $<$8.3$\times$10$^{-14}$ \\

\noalign{\smallskip}
\hline
\hline
\noalign{\smallskip}
\end{tabular}
\end{center}
\end{table*}


For four objects in our sample (see Table 4), we have X-ray 
observations acquired with the X-ray Telescope (XRT, 0.2--10 keV, Burrows 
et al. 2005) on board the \emph{Swift} satellite (Gehrels et al. 2004). 
XRT data reduction was performed using the XRTDAS standard data pipeline 
package ({\sc xrtpipeline} v. 0.12.4), in order to produce screened event 
files. All data were extracted only in the Photon Counting (PC) mode (Hill 
et al. 2004), adopting the standard grade filtering (0--12 for PC) 
according to the XRT nomenclature.

Events for spectral analysis were extracted within a circular region of 
radius 20$''$, centered on the source position, which encloses about 90\% 
of the PSF at 1.5 keV (see Moretti et al. 2004).

The background was taken from various source-free regions close to the 
X-ray source of interest, using circular regions with different radii in 
order to ensure an evenly sampled background. In all cases, the spectra 
were extracted from the corresponding event files using the {\sc XSELECT} 
software and binned using {\sc grppha} in an appropriate way, so that the 
$\chi^{2}$ statistic could be applied. We used version v.011 of the 
response matrices and create individual ancillary response files 
\textit{arf} using {\sc xrtmkarf v. 0.5.6}.

The data have been fitted using an absorbed powerlaw model; due to the 
poor statistical quality of the X-ray data, we have fixed the photon index 
to 1.8 in order to evaluate the presence of absorption and the 2--10 keV 
flux.

For four sources in the sample (see Table 5) optical spectroscopy of the 
proposed counterparts was obtained from data collected at the 1.5-m 
telescope of the Cerro Tololo Interamerican Observatory (CTIO), Chile, at 
the 2.1-m telescope of the Observatorio Astron\'omico Nacional in San 
Pedro M\'artir (SPM), M\'exico, and from the Six-degree Field Galaxy 
Survey\footnote{http://www.aao.gov.au/local/www/6df/} (6dFGS) archive 
(Jones et al. 2004) containing spectra acquired with the 4-m 
Anglo-Australian Telescope (AAT) in Siding Spring, Australia. Table 5 
reports the log of these observations.


\begin{table*}
\caption[]{Log of the spectroscopic observations presented in this paper.}
\begin{center}
\begin{tabular}{llcccc}
\noalign{\smallskip}
\hline
\hline
\noalign{\smallskip}
\multicolumn{1}{c}{Object} & \multicolumn{1}{c}{Date} & Telescope & Exp. 
start & Disp. & Exposure \\
 & & + instrument & time (UT) & (\AA/pix) & time (s) \\
\noalign{\smallskip}
\hline
\noalign{\smallskip}

LCSB L0289O & 13 Jan 2005 & AAT+6dF & 12:15 & 1.6 & 1200+600 \\

USNO-A2.0 0450\_06519994 & 18 Jan 2010 & CTIO 1.5m + RC Spec. & 03:54 & 
5.7 & 2$\times$1200 \\

UGC 11802 & 15 Sep 2009 & SPM 2.1m + B\&C Spec. & 04:37 & 4.0 & 
2$\times$1800 \\

UGC 11806 & 15 Sep 2009 & SPM 2.1m + B\&C Spec. & 05:59 & 4.0 & 
2$\times$1800 \\

\noalign{\smallskip}
\hline
\noalign{\smallskip}
\end{tabular}
\end{center}
\end{table*}


The spectroscopic data acquired at these telescopes were optimally 
extracted (Horne 1986) and reduced following standard procedures using 
IRAF\footnote{IRAF is the Image Reduction and Analysis Facility made 
available to the astronomical community by the National Optical Astronomy 
Observatories, which are operated by AURA, Inc., under contract with the 
US National Science Foundation. It is available at http://iraf. noao.edu}. 
Calibration frames (flat fields and bias) were taken on the day preceeding 
or following the observing night. The wavelength calibration was performed 
using lamp data acquired soon after each on-target spectroscopic 
acquisiton; the uncertainty in this calibration was $\sim$0.5 \AA~in all 
cases, according to our checks made using the positions of background 
night sky lines. Flux calibration was performed using catalogued 
spectrophotometric standards.

As mentioned above an additional spectrum, for 2MASX J05580231--1255477, 
was retrieved from the 6dFGS archive. Since this archive provides spectra 
that are not flux-calibrated, we used the optical photometric information 
in Jones et al. (2005) to calibrate the 6dFGS spectrum presented in this 
work.

The flux calibration was obtained by normalizing the count spectrum 
and by multiplying it by a cubic spline constructed with the fluxes 
extracted from the $BRI$ optical magnitudes available for the source using 
the conversion formulae of Fukugita et al. (1995).

The results of these X-ray and optical follow up observations are 
presented in Table 4 and Table 6.

\section{Results}

In the following, the results of all available archival information 
gathered on each individual source are discussed together with the X-ray 
and optical data when available .

As for the X-ray data, we note that in 3 cases (IGR J08190-3835, IGR 
J17520-6018 and IGR J21441+4640) the X-ray measurements confirm the 
proposed counterpart; in the case of the galaxy pair only one of the two 
objects, UGC 11806, was detected while an upper limit was set on the 
other, UGC 11802. Only in one case (IGR J05583--1257) no X-ray source is 
detected within the {\it INTEGRAL} error circle (see Section on this 
source) and an upper limit is reported for the proposed counterpart.

For the optical spectra categorization, as they all refer to extragalactic 
sources (see below), we used the criteria of Veilleux \& Osterbrock (1987) 
and the line ratio diagnostics of both Ho et al. (1993, 1997) and 
Kauffmann et al. (2003) which are generally used for emission-line AGN 
classification. The spectra of the galaxies shown here were not corrected 
for starlight contamination (see, e.g., Ho et al. 1993, 1997) because of 
the limited S/N and spectral resolution. In this case, we do not expect 
this to affect any of our main results and conclusions.

\begin{figure*}
\hspace{-.1cm}
\centering{\mbox{\psfig{file=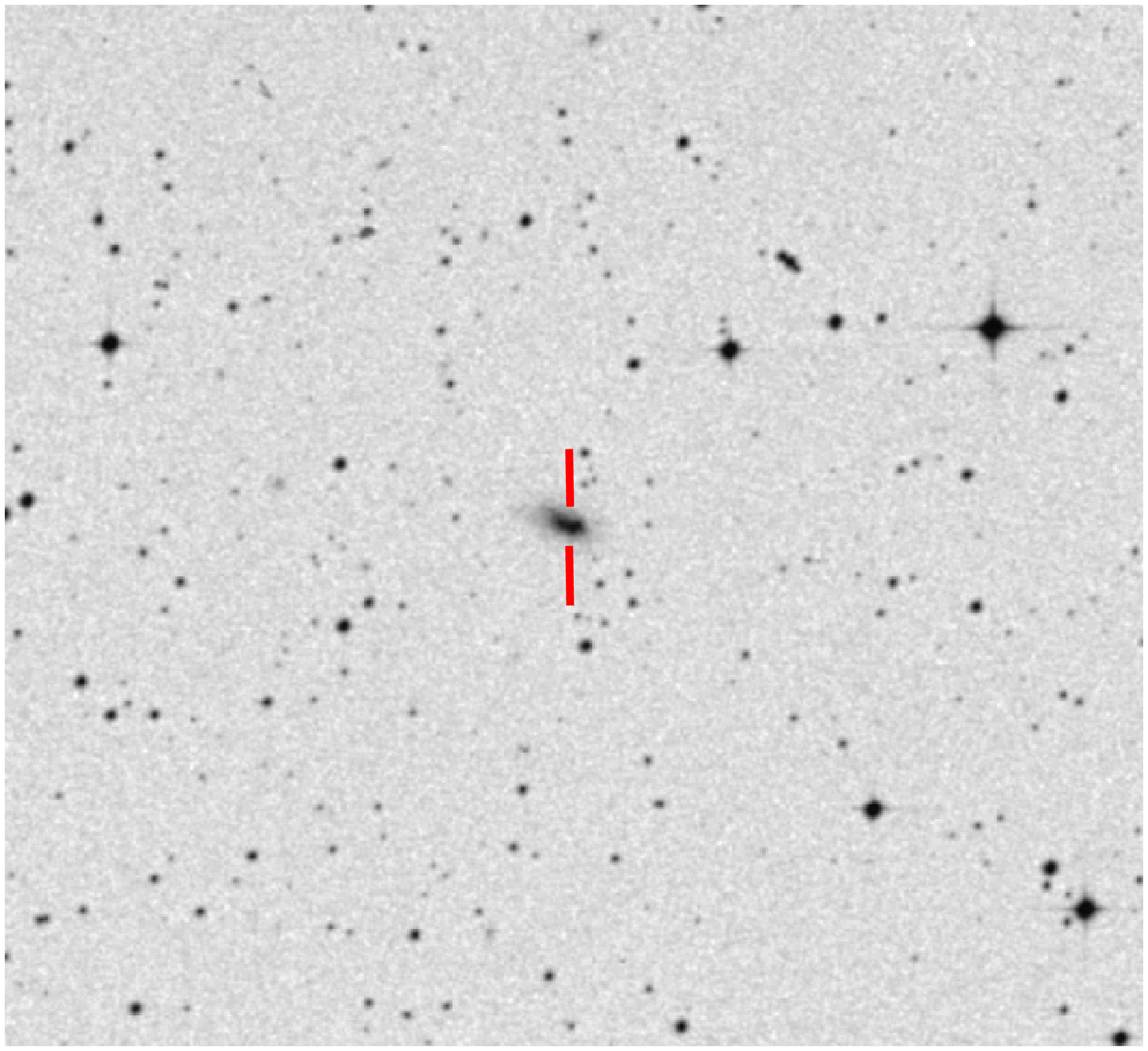,width=5.7cm}}}
\centering{\mbox{\psfig{file=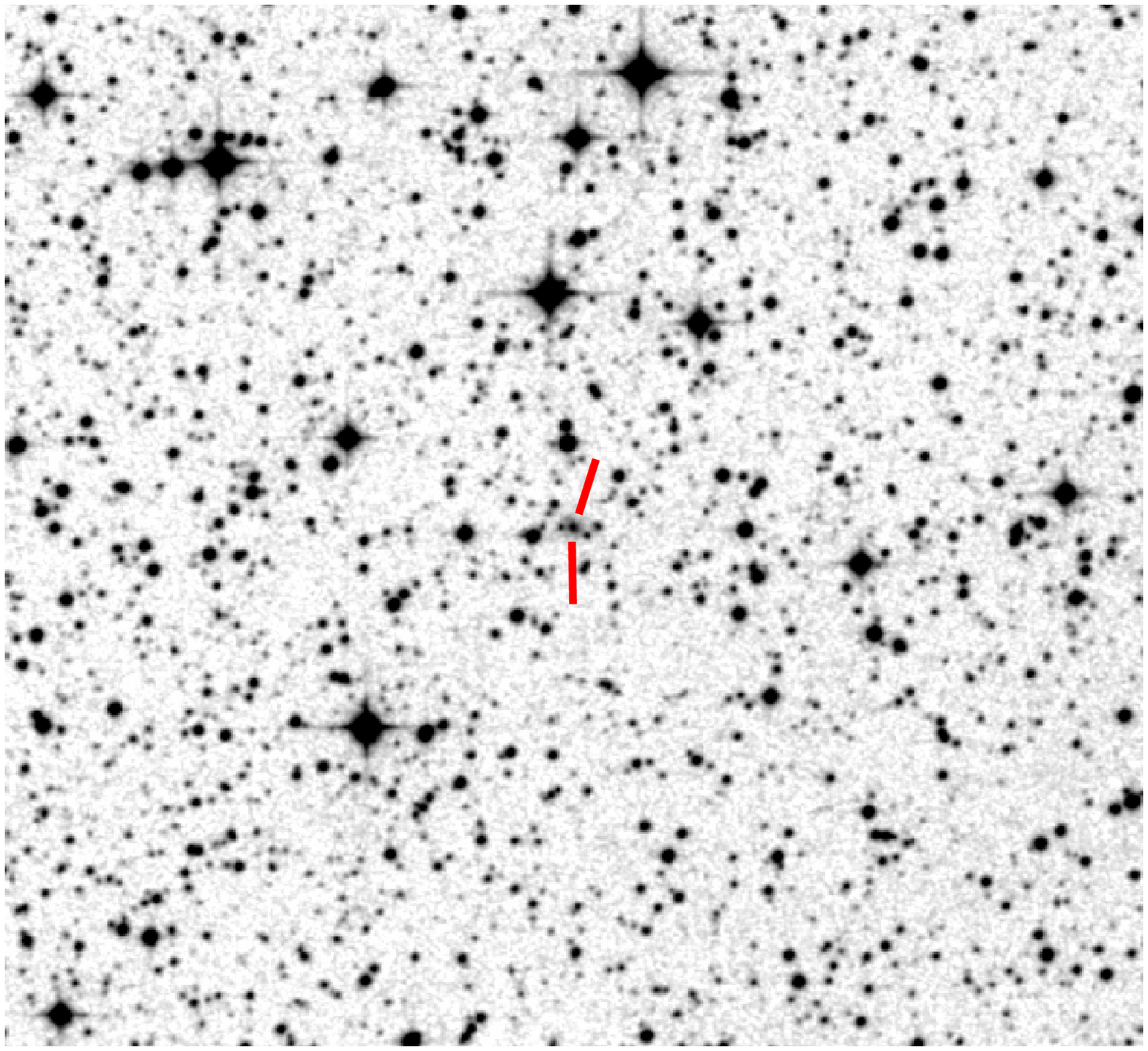,width=5.6cm}}}
\centering{\mbox{\psfig{file=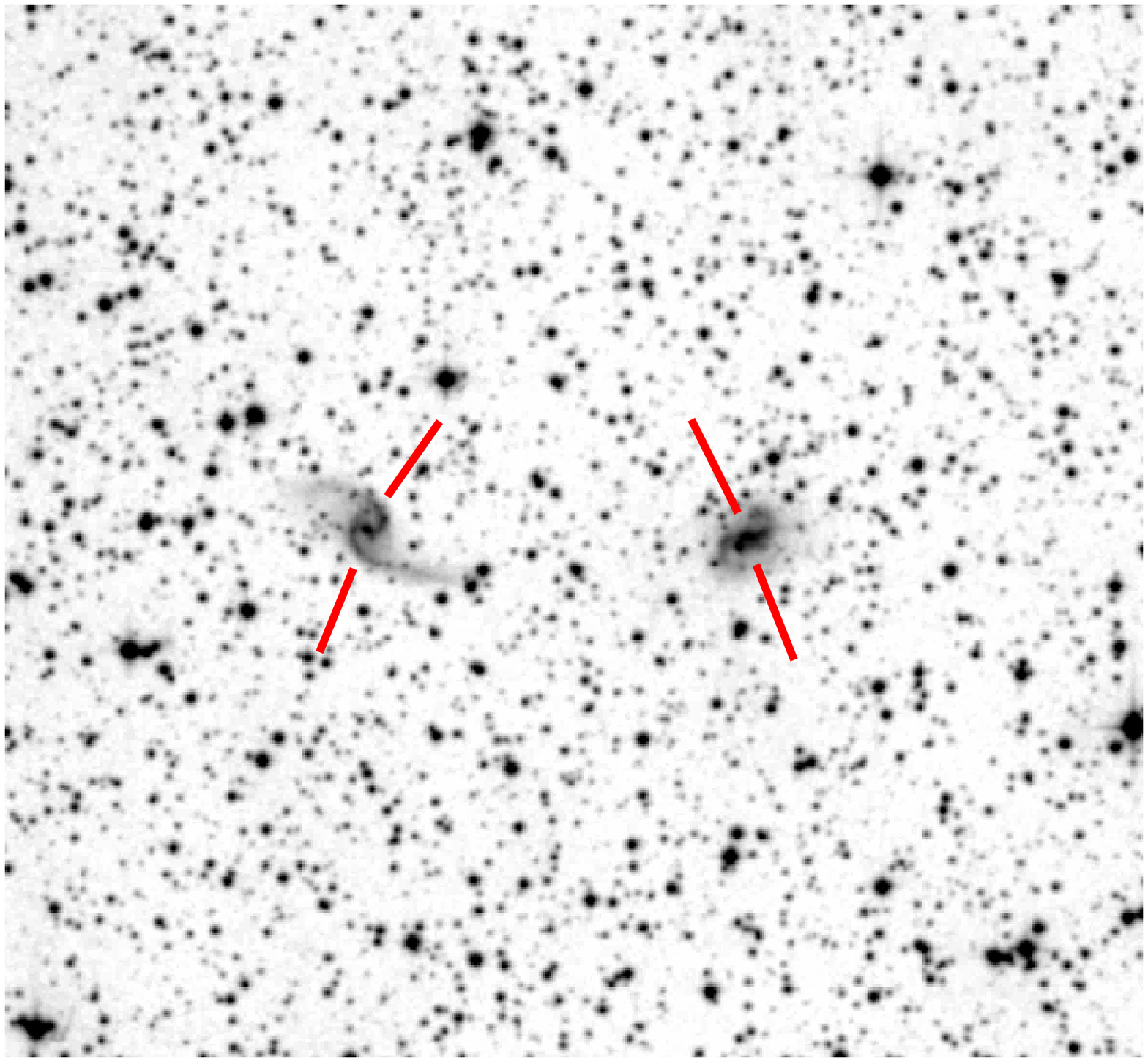,width=5.6cm}}}

\caption{Optical images of the fields of 3 of the {\it INTEGRAL} hard 
X-ray sources selected in this paper for optical spectroscopic 
follow-up: IGR J05583--1257 (left panel), IGR J08190--3835 (central panel) 
and IGR J21441+4640 (right panel). 
The proposed optical counterparts are indicated with tick marks (see text). 
Field sizes are 10$'$$\times$10$'$ and are extracted from the DSS-II-Red 
survey. In all cases, north is up and east to the left.}
\end{figure*}

Of the four objects for which we have obtained optical spectra, one was 
found to be a type 2 AGN (IGR J08190--3835), two (i.e. those belonging to 
the galaxy pair) were classified as LINERs and one shows the features 
typical of a starburst galaxy (IGR J05583--1257).

In the following, we consider a cosmology with $H_{\rm 0}$ = 71 km 
s$^{-1}$ Mpc$^{-1}$, $\Omega_{\Lambda}$ = 0.73, and $\Omega_{\rm m}$ = 
0.27; the luminosity distances of the extragalactic objects reported in 
this paper were computed for these parameters using the Cosmology 
Calculator of Wright (2006).

\bigskip

\noindent
{\bf IGR J00556+7708}

\noindent
This source, located at high Galactic latitude, is detected by IBIS in a 
Revolution Sequence suggesting that it might be variable on long 
timescales. As said above, in this case the radio and NIR positions are 
significantly different and the objects are located on the left edge of 
the IBIS error box (Fig. 1, top-left panel). The radio object is detected 
at 20 and 92 cm and has a steep spectrum with index $\alpha$=0.79, typical 
of a radio loud AGN; no optical counterpart is found within the radio 
source positional uncertainty. The NIR source is also listed in the 
USNO-A2.0 
catalogue\footnote{http://www.nofs.navy.mil/projects/pmm/USNOSA2doc.html} 
with name USNO-A2.0 1650\_00208794 and optical magnitudes R$\sim$17.4 and 
B$\sim$19.5; the NIR colour indices are typical of an active galaxy 
(Kouzuma \& Yamaoka, 2010).

Overall, we conclude that the radio and NIR objects are both likely AGN 
and, as such, are both possible counterparts of the IBIS detection.

\bigskip

\noindent
{\bf IGR J03103+5706}

\noindent
This object is also likely variable as it was detected by IBIS 
during just one revolution (3 days). The radio/NIR counterpart suggested 
in this work is located within the IBIS error circle (Fig. 1, top-right 
panel). It is detected in radio at 20 and 92 cm and has $\alpha$=0.50, 
which is characteristic of a radio flat spectrum galaxy. The source has no 
optical counterpart in USNO-B1.0 (Monet et al. 2003), suggestive of a 
highly reddened/absorbed object. Its NIR photometry, as well as its 
location on the Galactic plane, are not typical of an AGN, so the nature 
of this object remains dubious.

\bigskip

\noindent
{\bf IGR J05583--1257}

\noindent 
This is one of the two persistent sources in the sample; the radio/NIR 
source is identified in NED with LCSB L0289O, a low brightness galaxy at z 
= 0.003. This source is fairly bright in the far IR being detected 
by IRAS at 60 $\mu$m with a flux of 0.6 Jy (IRAS, 1988). The available 
radio data (Fig. 1, bottom left panel) for this source only consist of a 
flux at 20 cm, thus no information on the spectral index can be obtained. 
The 6dFGS optical spectrum (Fig. 4, first panel from top; see also 
DSS-II-red image in Fig. 3, left panel) shows several permitted and 
forbidden narrow emission lines at the above redshift; their ratios (see 
also Table 6) indicate that this is a starburst galaxy with no noticeable 
nuclear activity.

There is no X-ray detection within the 90\% IBIS error box, but at the 
border of the 99\% positional uncertainty we find a detection at low 
significance level, 2.6$\sigma$. This source is located at RA(J2000) = 05h 
57m 59.9s, Dec(J2000) = --12$^{\circ}$ 51$'$ 33.0$''$ (6$''$ uncertainty). 
The X-ray flux is $7.3 \times 10^{-14}$ erg cm$^{-2}$ s$^{-1}$, with 
photon index fixed to 1.8. At this position no optical or IR or radio 
source is found. On the other hand, we do not find any X-ray emission at 
the location of LCSB 0289O. The 2--10 keV flux upper limit provides an 
indication on the source luminosity which is $<$ 10$^{39}$ erg s$^{-1}$, 
i.e. quite low for an AGN, unless the source is extremely variable (but 
this is not evident in the IBIS data), or extremely absorbed (thus masking 
a Compton thick AGN in a starburst galaxy).

The persistent nature of the source, the stringent upper limit in X-rays 
and the optical spectrum suggest that this is probably not the counterpart 
of the IBIS source, which is either a spurious detection or has a 
different association than LCSB 0289O.

\bigskip

\noindent
{\bf IGR J08190--3835}

\noindent
This is the second persistent source in our sample. Within the IBIS 
uncertainty, XRT detects only one source with a statistical significance 
of $5.8\sigma$ in the energy range 0.3--10 keV.  This object is 
positionally coincident with both 2MASX and NVSS sources (Fig. 1, 
bottom-right panel). The radio analysis provides a flux of $\sim$3 mJy at 
20 cm. The source is fairly bright at NIR frequencies (see Table 2) and it 
is also detected in the optical band, where is listed in the USNO-A2.0 
catalogue with name USNO-A2.0 0450\_06519994 (see Table 5 and Table 6) and 
magnitudes R$\sim$14.6 and B$\sim$16.8. From a comparison with the NIR 
magnitudes in Table 2, the R--K colour is therefore $\sim$ 4, which 
suggests a red or obscured galaxy (Fig. 3, central panel and Table 6).

The X-ray data analysis is described by an absorbed power law with photon 
index fixed to 1.8 and an observed 2--10 keV flux of $1.5 \times 10^{-12}$ 
erg cm$^{-2}$ s$^{-1}$; the intrinsic column density is N$_{\rm H}(intr) 
\sim 1.4 \times 10^{23}$ cm$^{-2}$, which exceedes the Galactic value of 
$9.6 \times 10^{21}$ cm$^{-2}$ (Kalberla et al. 2005). Optical 
spectroscopy (Fig. 4, second panel from top) shows that the source 
displays H$_{\alpha}$, [N {\sc ii}] and [S {\sc ii}] narrow emission lines 
at redshift $z=0.009\pm0.001$ superimposed on a very reddened continuum. 
The flux ratios among these emission features suggest that the object is a 
Type 2 AGN.

Overall, we conclude that the proposed 2MASX/NVSS/XRT source is the actual 
counterpart of the IBIS-detected object and it is a Narrow Line (obscured) 
AGN.


\bigskip

\noindent
{\bf IGR J17219--1509}

\noindent
The radio/NIR source is located at the edge of the IBIS error circle (Fig. 
2, top-left panel); it is very weak in radio, being close to the flux 
limit of the NVSS map ($\sim$5 mJy). Very little is known about this 
source except that is strongly variable in the IBIS waveband, being 
detected only in one revolution (R403, i.e. for a few days) and showing a 
large bursticity factor of 4 (see Bird et al. 2010 for details). The 
source has an optical magnitude of 18.2 in R; the NIR colours, as well as 
its location above the Galaxy plane, are compatible with the source being 
a nearby AGN (Kouzuma \& Yamaoka, 2010).

We conclude that this is an extragalactic object, most likely an AGN, on 
the basis of the above considerations and it is a good association for 
the {\it INTEGRAL} source.

\bigskip

\noindent
{\bf IGR J17520--6018}

\noindent
This source can also be considered as variable by Bird et al. 
(2010). An X-ray source is well detected at $13.6\sigma$ confidence level 
in the range 0.3--10 keV by XRT within the IBIS uncertainty and is located 
at RA(J2000) = 17h 51m 55.9s and Dec(J2000) = --60$^{\circ}$ 19$'$ 44$''$ 
(4$''$ uncertainty). It coincides with both the 2MASS extended object and 
the radio source reported in the SUMSS (Fig. 2, top-right panel). The 
source has only one detection in radio at 36 cm. Besides being detected in 
the NIR, the galaxy is also listed in the USNO-B1 catalogue with 
optical B and R magnitudes of 15.5 and 14.7, respectively. The X-ray data 
analysis provides an absorbed power law spectrum with fixed photon index 
of 1.8 and an observed 2--10 keV flux of $2.6 \times 10^{-12}$ erg 
cm$^{-2}$ s$^{-1}$; the intrinsic column density is N$_{\rm H}(intr) = 1.3 
\times 10^{23}$ cm$^{-2}$ which exceedes the Galactic value (see Table 4). 
This source is also reported in the BAT 58--Month catalogue\footnote{ 
http://swift.gsfc.nasa.gov/docs/swift/results/bs58mon/} as SWIFT 
J1751.8--6019, with flux 1.7$\times$10$^{-11}$ erg cm$^{-2}$ s$^{-1}$ in 
the range 14--195 keV.

Based on all the above information, we conclude that this galaxy is the 
likely counterpart of the IBIS/BAT source and, on the basis of the 
detected X-ray absorption, we further suggest that it is a type 2 AGN.

\bigskip

\noindent
{\bf IGR J21268+6203}

\noindent
This IBIS source is also highly variable being reported with a 
high bursticity factor by Bird et al. (2010); it is also likely an 
extragalactic source being located above the Galactic plane. As said in 
Sect. 2.1, in this case the radio and NIR positions are significantly 
different (Fig. 2, bottom-left panel). The NIR object (Table 2) has an 
optical counterpart in USNO-B1.0 1520\_0331686 listed in USNO-B1.0 
catalogue with magnitudes B$\sim$20.3, R$\sim$17.5 and I$\sim$16.4; no 
optical association was found for the radio object. The NIR colours are 
also in this case typical of a nearby AGN (Kouzuma \& Yamaoka, 2010). The 
radio data available provide a spectral index $\alpha$=--0.3 between 20 
and 92 cm, indicative of a probable GHz-Peaked-Spectrum (GPS) radio 
source, i.e. a source characterised by a convex radio spectrum peaking 
near 1 GHz (Stanghellini 2006; Lister 2003; O'Dea 1998). GPS are compact 
powerful young radio galaxies that reside in gas-rich environments at the 
center of active galaxies.

Thus, the observational evidence suggests that the radio and NIR objects 
are both likely AGN and, as such, are good candidates for an association 
with the {\it INTEGRAL} source.

\begin{figure}
\vspace{-.3cm}
\mbox{\psfig{file=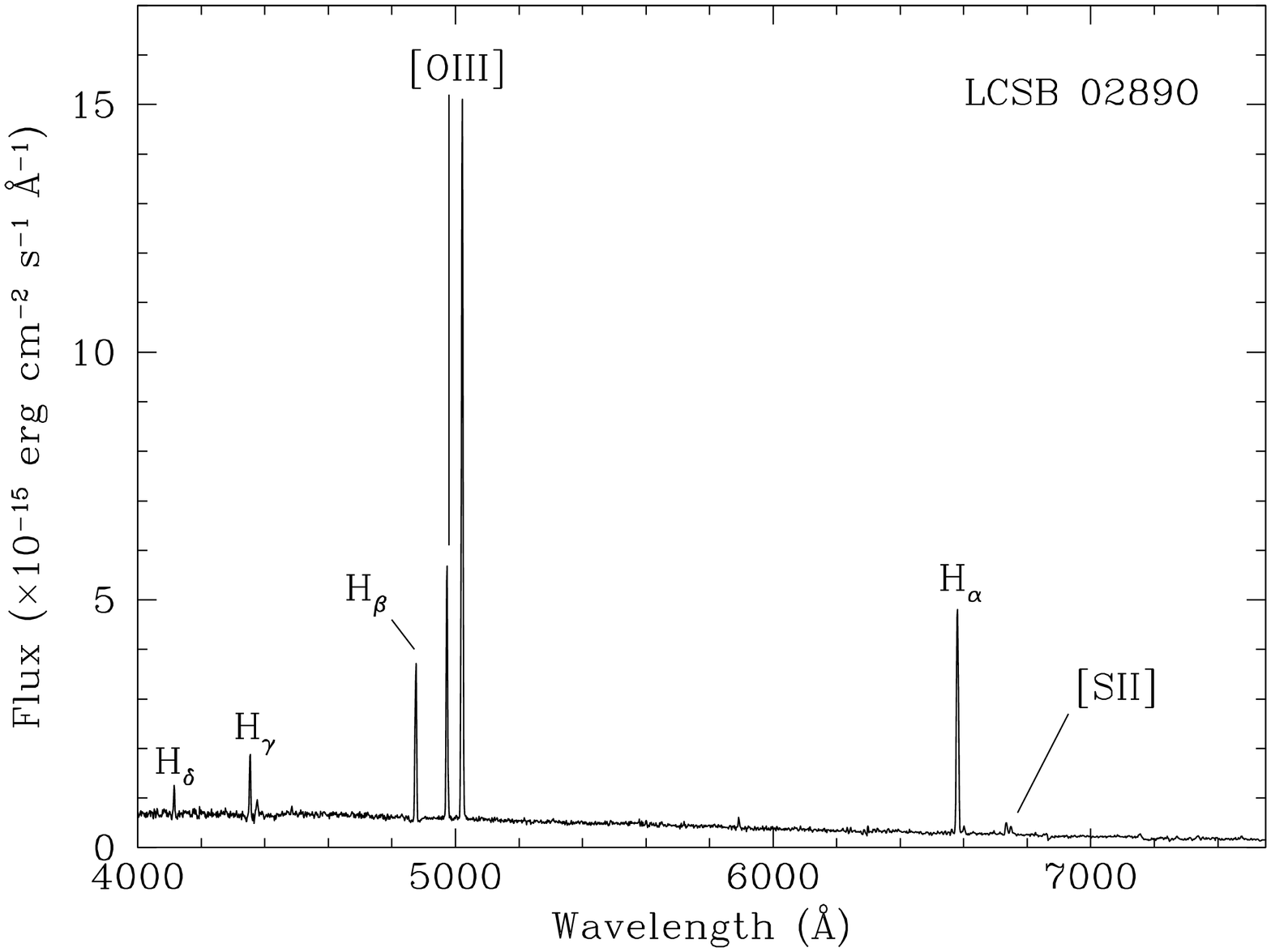,width=5.8cm,angle=270}}
\mbox{\psfig{file=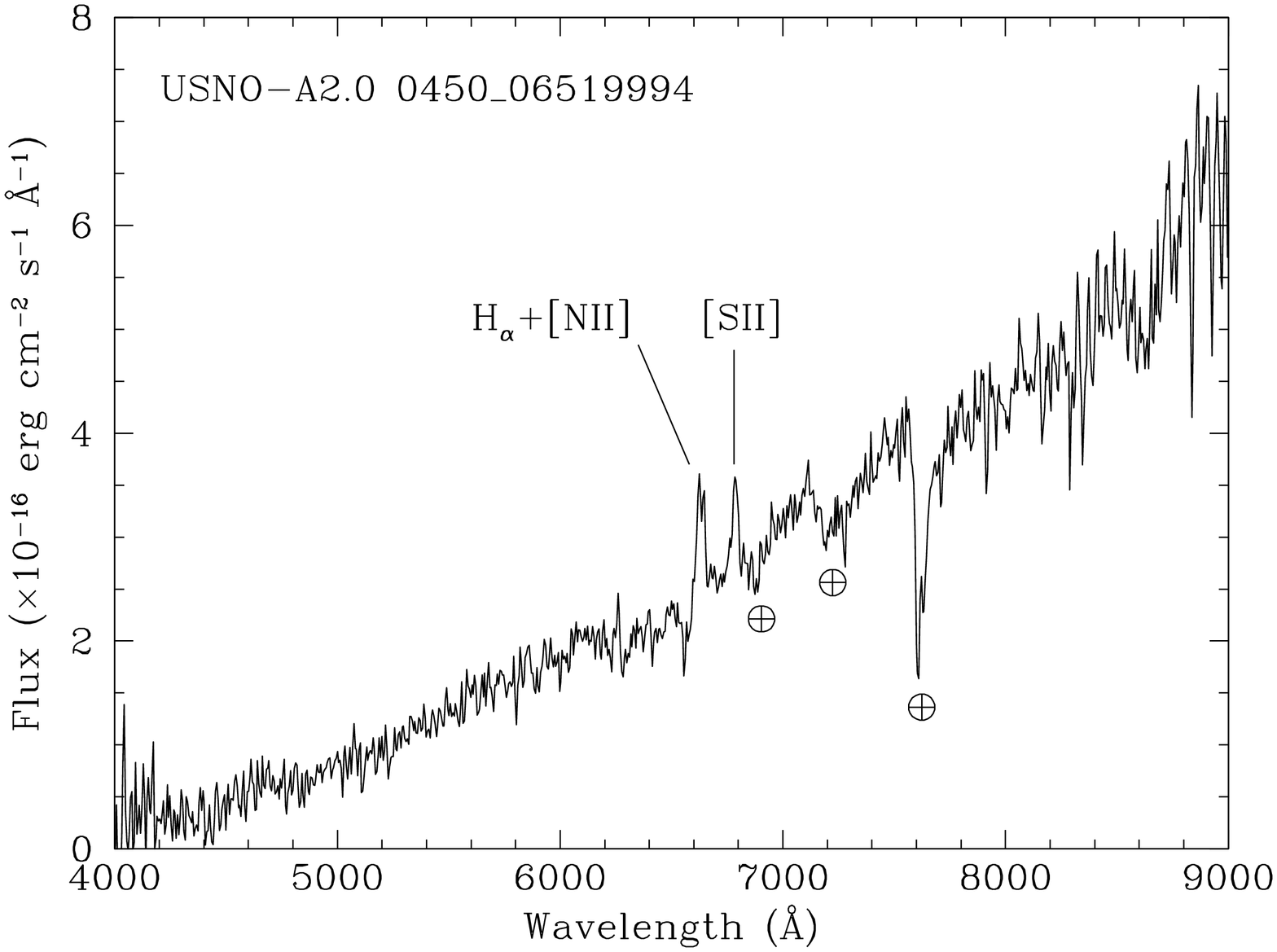,width=5.8cm,angle=270}}

\vspace{-.7cm}
\mbox{\psfig{file=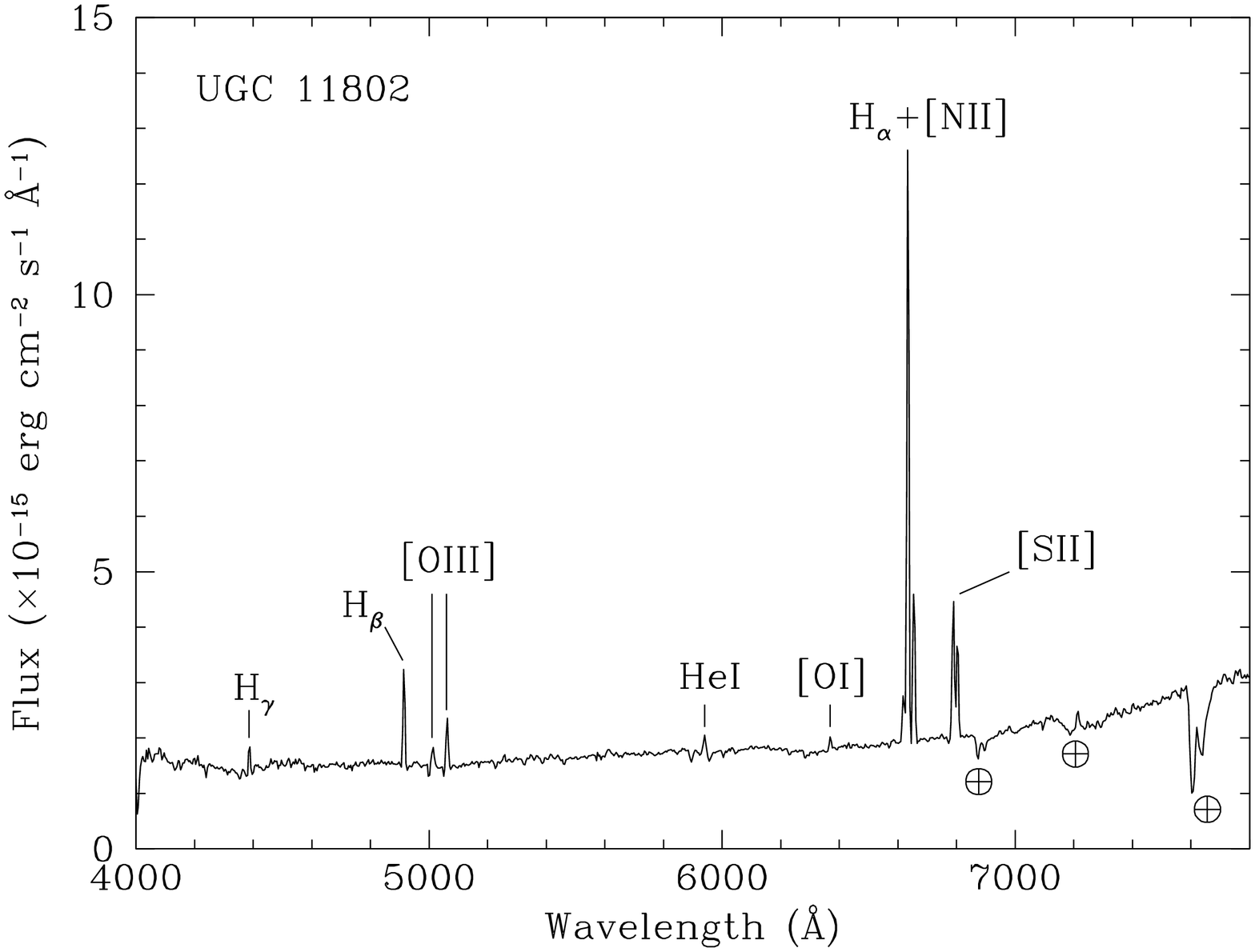,width=5.8cm,angle=270}}
\mbox{\psfig{file=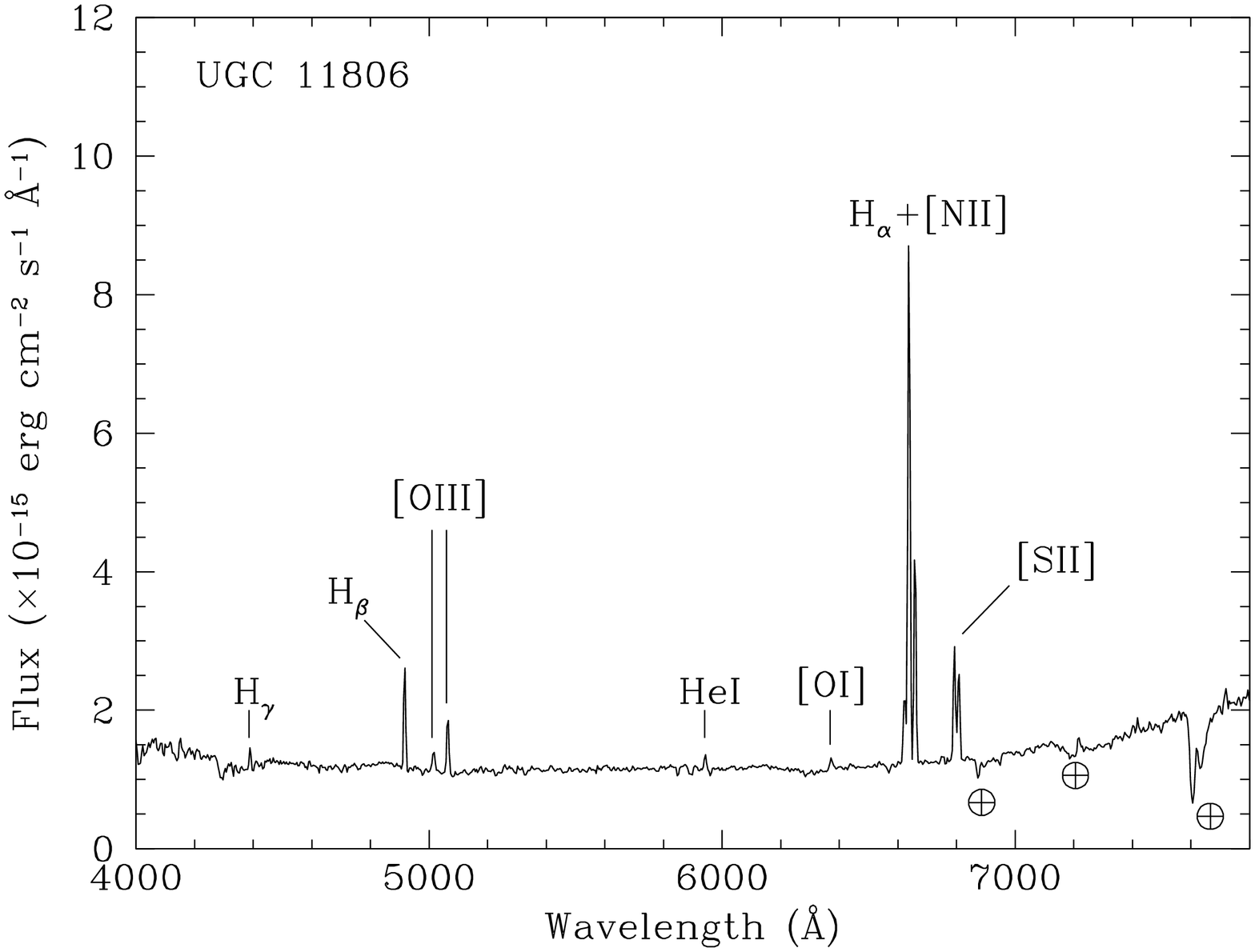,width=5.8cm,angle=270}}

\vspace{-.5cm}
\caption{Spectra (not corrected for the intervening Galactic absorption) 
of the possible optical counterparts of some of the {\it INTEGRAL} sources 
presented in this paper (IGR J05583--1257, IGR J08190--3835 and IGR 
J21441+4640; see text). For each spectrum, the main spectral features are 
labeled. The symbol $\oplus$ indicates atmospheric telluric absorption 
bands.}
\end{figure}


\begin{table}
\caption[]{Fluxes (in units of 10$^{-14}$ erg cm$^{-2}$ s$^{-1}$) of 
the main emission lines detected in the spectra of the objects reported 
in Fig. 4. The correction for the Galactic reddening was computed 
a color excess $E(B-V)$ = 0.479, 1.54, 0.404 and 0.403 mag for LCSB 
L0289O, USNO-A2.0 0450\_06519994, UGC 11802 and UGC 11806, respectively 
(from Schlegel et al. 1998). Uncertainties and upper limits for the 
fluxes are reported at 1$\sigma$ and 3$\sigma$ confidence levels, 
respectively.}
\begin{center}
\begin{tabular}{lrr}
\noalign{\smallskip}
\hline
\hline
\noalign{\smallskip}
\multicolumn{1}{c}{Line} & \multicolumn{1}{c}{Observed flux} & 
\multicolumn{1}{c}{Corrected flux} \\
\noalign{\smallskip}
\hline
\noalign{\smallskip}

\multicolumn{3}{c}{LCSB L0289O ($z=0.003$)} \\

H$_\beta$                       &  17.5$\pm$0.9  &  87$\pm$4    \\
$[$O {\sc iii}$]$ $\lambda$5007 &  92$\pm$3      & 428$\pm$13   \\
$[$O {\sc i}$]$ $\lambda$6300   &  $<$0.5        &  $<$2.2      \\
H$_\alpha$                      &  34.2$\pm$1.0  & 105$\pm$3    \\
$[$N {\sc ii}$]$ $\lambda$6583  &  1.0$\pm$0.3   &  3.3$\pm$1.0 \\
$[$S {\sc ii}$]$ $\lambda$6716  &  1.8$\pm$0.3   &  5.5$\pm$0.9 \\
$[$S {\sc ii}$]$ $\lambda$6731  &  1.2$\pm$0.2   &  3.6$\pm$0.6 \\

& & \\

\multicolumn{3}{c}{USNO-A2.0 0450\_06519994 ($z=0.009$)} \\

H$_\beta$                       &  $<$0.07 &  $<$9 \\
$[$O {\sc iii}$]$ $\lambda$5007 &  $<$0.07 &  $<$9 \\
$[$O {\sc i}$]$ $\lambda$6300   &  $<$0.17 &  $<$7  \\
H$_\alpha$                      & 0.32$\pm$0.08 &  11$\pm$3 \\
$[$N {\sc ii}$]$ $\lambda$6583  & 0.12$\pm$0.03 &  4.1$\pm$1.0 \\
$[$S {\sc ii}$]^*$              & 0.26$\pm$0.04 &  8.3$\pm$1.2 \\

& & \\

\multicolumn{3}{c}{UGC 11802 ($z=0.0105$)} \\

H$_\beta$                       &  1.41$\pm$0.14 &  5.4$\pm$0.5 \\
$[$O {\sc iii}$]$ $\lambda$5007 &  0.76$\pm$0.11 &  2.7$\pm$0.4 \\
$[$O {\sc i}$]$ $\lambda$6300   &  0.24$\pm$0.06 &  0.55$\pm$0.14 \\
H$_\alpha$                      &  9.5$\pm$0.3   & 24.1$\pm$0.7 \\
$[$N {\sc ii}$]$ $\lambda$6583  &  2.47$\pm$0.12 &  6.4$\pm$0.3 \\
$[$S {\sc ii}$]$ $\lambda$6716  &  2.26$\pm$0.16 &  5.6$\pm$0.4 \\
$[$S {\sc ii}$]$ $\lambda$6731  &  1.58$\pm$0.11 &  3.9$\pm$0.3 \\

& & \\

\multicolumn{3}{c}{UGC 11806 ($z=0.0110$)} \\

H$_\beta$                       &  1.22$\pm$0.12 &  4.8$\pm$0.5 \\
$[$O {\sc iii}$]$ $\lambda$5007 &  0.76$\pm$0.08 &  2.8$\pm$0.3 \\
$[$O {\sc i}$]$ $\lambda$6300   &  0.21$\pm$0.05 &  0.70$\pm$0.18 \\
H$_\alpha$                      &   7.0$\pm$0.2  & 17.7$\pm$0.5 \\
$[$N {\sc ii}$]$ $\lambda$6583  &  2.88$\pm$0.14 &  7.2$\pm$0.4 \\
$[$S {\sc ii}$]$ $\lambda$6716  &  1.53$\pm$0.11 &  3.7$\pm$0.3 \\
$[$S {\sc ii}$]$ $\lambda$6731  &  1.24$\pm$0.09 &  3.1$\pm$0.3 \\

\noalign{\smallskip}
\hline
\noalign{\smallskip}
\multicolumn{3}{l}{$^*$: The doublet is blended due to the low} \\
\multicolumn{3}{l}{spectral S/N and resolution.} \\
\noalign{\smallskip}
\hline
\noalign{\smallskip}
\end{tabular}
\end{center}
\end{table}


\bigskip

\noindent
{\bf IGR J21441+4640}

\noindent 
This also is a strongly variable source in IBIS (Bird et al. 2010) 
with a bursticity larger than 4. Within the IBIS positional uncertainty of 
this source there are two galaxies (see Fig. 3, right panel), UGC 11802 
and UGC 11806, at the same redshift ($z$=0.011), which form the galaxy 
pair KPG559. Both galaxies are detected at radio frequencies in the NVSS 
catalogue (NVSS J214354+463705, NVSS J214413+463718) and are also listed 
in the 2MASS extended catalogue (2MASX J21435408+4637048, 2MASX 
J21441345+4637169) (see Fig. 2, bottom-right panel). The radio data 
analysis provides a 20 cm flux for both objects, which are also detected 
at 92 cm and have spectral indices $\alpha$=0.38 and $\alpha$=0.51, 
respectively; both radio power and spectral index are typical of low 
luminosity AGN. The optical B (R) magnitudes of the two objects are 14.5 
(9.5) and 14.7 (12.9). Both are detected with AKARI in the far IR, 
from 90 to 160 $\mu$m at the few Jansky flux level (Murakami et al. 2007). 
Within the IBIS uncertainty, XRT finds only one galaxy (UGC 11806) of the 
pair, with a low statistical significance of $2.5\sigma$ in the range 
0.3--10 keV. The XRT spectrum can be described by an unabsorbed power law 
with fixed photon index of 1.8 and an observed 2--10 keV flux of $1.2 
\times 10^{-13}$ erg cm$^{-2}$ s$^{-1}$. The upper limit of the X-ray flux 
in the 2--10 keV band for the companion galaxy UGC11802 is $8.3\times 
10^{-14}$ erg cm$^{-2}$ s$^{-1}$.

Optical spectroscopy (Fig. 4, lower panel) indicates that UGC 11806 is a 
narrow emission-line galaxy with flat continuum and prominent Balmer, [N 
{\sc ii}], [O {\sc iii}] and [S {\sc ii}] lines at a redshift consistent 
with the one found in the literature. Emission line ratios suggest that 
this is a transition object, that is, a LINER (Heckman 1980) with a 
possible contamination from an underlying starburst event. The optical 
spectrum of UGC 11802 (Fig. 4, third panel from top) indicates instead 
that this galaxy is a starburst, with no indication of AGN activity (which 
explains the absence of detectable X-ray emission from its nucleus).

The low significance detection in X-rays of UGC 11806 may be due to the 
variable nature of the source (Bird et al. 2010) rather than to the high 
absorption, which is not readily apparent from the optical spectrum. 
To this aim we can infer the reddening local to the source by considering 
an intrinsic H$_\alpha$/H$_\beta$ line ratio of 2.86 (Osterbrock 1989). 
The corresponding color excess, obtained by comparing the intrinsic line 
ratio with the measured one by applying the Galactic extinction law of 
Cardelli et al. (1989), is $E(B-V)$ = 0.25 mag. This, using the formula of 
Predehl \& Schmitt (1995), corresponds to a hydrogen column density 
$N_{\rm H}$ = 1.4$\times$10$^{21}$ cm$^{-2}$ local to the AGN. 


Overall we conclude that UGC 11806 is the counterpart of the IBIS source; 
it is probably a low luminosity and highly variable AGN of LINER type.


\section{Conclusion}

The basic idea of this work is to propose a way whereby AGN can be 
easily found out among a set of unidentified objects detected in hard 
X-ray surveys. The method, which consists of two consecutive steps of 
cross-correlations between hard X-ray objects and IR/radio catalogues, is 
tested here for the unidentified sources contained in the 4th IBIS survey 
catalogue.

Following this procedure, we first used the 2MASS extended catalogue 
to identify galaxies in the IBIS error circle and then we extracted those 
which were also radio emitters, in order to isolate AGN candidates by 
means of NVSS and SUMSS radio catalogues. As a result we obtained a set 
of 8 objects for which we performed a more in-depth study, and in some 
cases optical and/or X-ray follow up observations, in order to verify 
their true association with the {\it INTEGRAL} source as well as their AGN 
nature and class.

The purpose of this work is not to search out for AGN using different 
selection criteria, but rather to confirm how a multiwavelength (radio, 
IR, optical, and X-ray) study of these 8 sources can be used to test the 
level of reliability and accuracy of the proposed method.

In three cases (IGR J08190--3835, IGR J17520--6018, IGR J21441+4640) we 
found the X-ray counterparts of the IBIS sources. The optical spectra 
obtained for two of these sources (IGR J08190--3835, IGR J21441+4640) 
allowed us to identify them as AGN belonging to the Type 2 and LINER 
class. The third one (IGR J17520--6018) is most likely a type 2 AGN on the 
basis of the high X-ray absorption measured. We further suggest that 3 
sources (IGR J00556+7708, IGR J17219--1509, IGR J21268+6203) are likely 
active galaxies on the basis of the radio spectra, NIR photometry and 
location above the Galactic plane: they are all likely associated with the 
IBIS objects. LCSB 0289O is instead a starburst galaxy, which is at most a 
very weak X-ray source and so unlikely to emit at high energies; we 
conclude that it is an improbable association of IGR J05583--1257. The 
nature of this {\it INTEGRAL} source is therefore still open. In only one 
case (IGR J03103+5706), we have not enough information for a clear 
classification of the radio/NIR source as an AGN; the nature of this {\it 
INTEGRAL} object remains doubious.

Overall, detailed information and follow up measurements confirm the 
goodness of our method in the search for AGNs among unidentified hard 
X-ray emitters. On the basis of this work we have classified in the 4th 
IBIS catalogue the sources discussed in this paper as AGNs and the same 
approach can be used to pinpoint AGN candidates among Swift/BAT or 
future Nustar (Harrison et al. 2010) unidentified objects.

\section*{Acknowledgments}

We thank Jos\'e Vel\'asquez for Service Mode observations at the CTIO 1.5m 
telescope, and Fred Walter for relaying the observing information to him.
We also thank the anonymous referee for useful remarks which helped us
to improve the quality of this paper. 
This research has made use of the ASI Science Data Center Multimission 
Archive; it also used the NASA Astrophysics Data System Abstract Service, 
the NASA/IPAC Extragalactic Database (NED), and the NASA/IPAC Infrared 
Science Archive, which are operated by the Jet Propulsion Laboratory, 
California Institute of Technology, under contract with the National 
Aeronautics and Space Administration. This publication made use of data 
products from the Two Micron All Sky Survey (2MASS), which is a joint 
project of the University of Massachusetts and the Infrared Processing and 
Analysis Center/California Institute of Technology, funded by the National 
Aeronautics and Space Administration and the National Science Foundation. 
This research has also made use of data extracted from the Six-degree 
Field Galaxy Survey and the Sloan Digitized Sky Survey archives; it has 
also made use of the SIMBAD database operated at CDS, Strasbourg, France, 
and of the HyperLeda catalog operated at the Observatoire de Lyon, France. 
The authors acknowledge the ASI and INAF financial support via grant No. 
I/008/07. LM is supported by the University of Padua through grant No. 
CPDR061795/06. VC is supported by the CONACYT research grant 54480-F 
(M\'exico). DM is supported by the Basal CATA PFB 06/09, and FONDAP Center 
for Astrophysics grant No. 15010003. GG acknowledges the support of 
Fondecyt regular 1085267, Fondo Gemini 32090009, and Fondo ALMA 31090009.
The authors acknowledge the ASI financial support via ASI--INAF contracts
I/033/10/0 and I/009/10/0.

\end{document}